\documentclass[aps,pra,floatfix,twocolumn,showpacs,amsmath,amssymb,letterpaper]{revtex4-2}
\usepackage{graphicx}
\usepackage{array}
\usepackage{bigstrut}
\usepackage{longtable}
\usepackage{rotating}
\usepackage{diagbox}
\usepackage[mode=text]{siunitx}
\usepackage{bm}
\usepackage{float}
\usepackage{lipsum}
\usepackage{warpcol}
\usepackage{color}
\usepackage{amsthm,amsmath}
\usepackage{mathrsfs}
\usepackage{appendix}
\usepackage{multirow}
\usepackage{threeparttable}
\usepackage{soul}
\usepackage[colorlinks=true, linkcolor=blue, citecolor=blue, urlcolor=blue]{hyperref}
\usepackage{nameref}
\usepackage{csquotes}
\usepackage{tabularx}
\usepackage{booktabs} 
\usepackage{dcolumn}
\allowdisplaybreaks

\newcolumntype{d}[1]{D{.}{.}{#1}}

\begin{document}



\title{Ionization energies for Rydberg $^4 \mathrm{He}$ ($1snp\,^{1,3}P$) states using the correlated B-spline basis function method}



\author{Jing Chi$^{1, 2}$}
\thanks{These authors contributed equally to this work.}
\author{Hao Fang$^{1}$}
\thanks{These authors contributed equally to this work.}
\author{Yong-Hui Zhang$^{1}$}
\author{Li-Yan Tang$^{1}$}
\email{lytang@apm.ac.cn}
\author{Ting-Yun Shi$^{1}$}
\email{tyshi@wipm.ac.cn}

\affiliation{
    $^{1}$
    Innovation Academy for Precision Measurement Science and Technology,
    Chinese Academy of Sciences, Wuhan 430071, China
}

\affiliation{
    $^{2}$University of Chinese Academy of Sciences, Beijing 100049, China
}


\date{\today}

\begin{abstract}

    We extend the correlated B-spline basis function (C-BSBF) method to high-precision calculations of the ionization energies of helium Rydberg $n^{1,3}P$ states ($n=24$--$35$). Using a unified basis set, we evaluate nonrelativistic energies, relativistic corrections of order $m\alpha^4$ (including finite-mass recoil), QED contributions of order $m\alpha^5$, and partial $m\alpha^6$ terms (singlet-triplet mixing, one- and two-loop radiative corrections). The remaining higher-order contributions are estimated via $1/n^3$ scaling. The resulting ionization energies achieve kHz-level accuracy and are in excellent agreement with independent Hylleraas calculations, thereby providing cross-validation between two distinct theoretical approaches. From these data, the quantum-defect parameters are determined and used to extrapolate the ionization energies up to $n=102$. Combining our Rydberg ionization energies with high-precision experimental $2S \rightarrow nP$ transition frequencies yields the ionization energies for the metastable $2^{1}S$ and $2^{3}S$ states as \num{960332040.533(10)}$_\mathrm{stat}(20)_ \mathrm{sys}$~MHz and \num{1152842742.7274(53)}$_\mathrm{stat}(25)_ \mathrm{sys}$~MHz, respectively. The C-BSBF result for the $2 \, ^1 S$ state is consistent with the experimental ionization energy obtained from Rydberg-series extrapolation, while for the $2 \, ^3 S$ state the difference is 0.019(10) MHz.

\end{abstract}

\pacs{31.30.J-, 31.15.-p, 31.15.ac}

\maketitle

\section{Introduction}

Helium is the simplest multi-electron atom and a crucial benchmark for atomic structure physics and quantum electrodynamics (QED). With advances in optical frequency comb metrology and Doppler-free spectroscopy techniques, transitions and ionization energies of low-lying states have been measured, achieving part-per-trillion (ppt) precision~\cite{Rooij_2011, Zheng_2017, Rengelink_2018, Clausen_4He_2025, Clausen_3He_2025, Werf_2025, WenJinLu_2025, Steinebach_2026}. Such experimental precision renders theory-experiment comparisons sensitive to previously negligible higher-order contributions and residual systematic effects. These include radiative QED, recoil, and finite nuclear-size effects~\cite{Pachucki_2017, Pachucki_2021, QiXiaoQiu_2025, Pachucki_2024, Pachucki_2026}, as well as residual systematics associated with Doppler shifts, ac-Stark shifts, and post-selection~\cite{Rengelink_2018, WenJinLu_2025, Steinebach_2026}. Precision spectroscopy therefore provides stringent tests of two-electron bound-state QED and enables the extraction of fundamental constants and nuclear properties, such as nuclear charge radii, from atomic spectra~\cite{Zheng_fine_2017, Werf_2025, Clausen_4He_2025, Steinebach_2026}.

Recently, Rydberg-series spectroscopy has pushed ionization energy determinations to higher precision and enabled stringent comparisons with theoretical predictions~\citep{Clausen_2021, Clausen_2023, Clausen_4He_2025, Clausen_3He_2025}. For $^4$He, the $2 \, ^3S_1$ ionization energy extracted from the $np$ Rydberg series differs from the \emph{ab initio} calculation by $0.4772 \pm 0.0523$~MHz, corresponding to a $9 \sigma$ discrepancy~\cite{Pachucki_2021, Clausen_4He_2025}. Discrepancies between experiment and theory have also been reported for several transitions involving low-lying states, most notably for the $2 \, ^3S_1$--$3 \, ^3D_1$ and $2 \, ^3P_0$--$3 \, ^3D_1$ transition frequencies, whose experimental values differ from the theoretical predictions by $6\sigma$ and $15\sigma$, respectively~\cite{Pachucki_2021, Yerokhin_2020, Pachucki_2011, Dorrer_1997, Luopeiling_2016}. Taken together, these comparisons indicate that, at the current experimental and theoretical accuracy, contributions beyond the present theoretical treatments and residual experimental systematics may still play a role, motivating further scrutiny on both sides.

\begin{figure}
    \centering
    \includegraphics[scale=0.33]{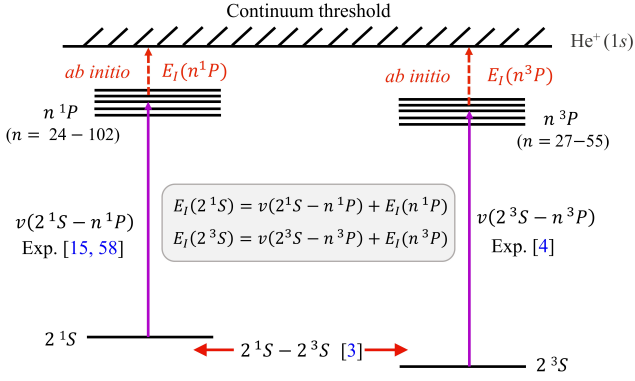}
    \caption{Scheme for determining the $2 \,^1 S$ and $2 \, ^3S$ ionization energies from measured Rydberg transitions and \emph{ab initio} calculations (not drawn to scale).}
    \label{fig:energy}
\end{figure}

High-$n$ Rydberg states are attractive because QED corrections scale approximately as $1/n^3$ and decrease with increasing $n$, so that higher-order QED contributions can often be neglected. This enables high-accuracy \emph{ab initio} calculations feasible in the Rydberg region, providing an alternative route to ionization limits without relying on empirical quantum defect theory (QDT). Nevertheless, achieving high accuracy for Rydberg states remains numerically challenging, since the accuracy of variational energy calculations decreases with increasing $n$. Consequently, most high-accuracy theoretical studies have been limited to low- and intermediate-lying states with principal quantum numbers up to $n \sim 10$~\cite{Drake_1992, Drake_1999, Yerokhin_2010, Aznabaev_2018, WuFangFei_2018, Drake_2023springer}. Very recently, Bondy \emph{et al.}~\cite{Bondy_2025} performed kHz-accuracy \emph{ab initio} calculations for the ionization energies of the $24 \, ^{1,3} P$ Rydberg states using ``triple'' Hylleraas basis sets, including relativistic and QED corrections. This opened a route to determining metastable-state ionization energies by combining precision measurements of transitions to the Rydberg series~\cite{Clausen_2021, Clausen_2023, Clausen_4He_2025, Clausen_3He_2025} with \emph{ab initio} ionization energies of the corresponding Rydberg states, as shown in Fig.~\ref{fig:energy}. On this basis, Drake \emph{et al.}~\cite{Drake_2026} extended the calculations for the $1snp \, ^{1,3}P$ states to $n = 35$, determined metastable-state ionization energies without using quantum-defect extrapolations, and confirmed the $9\sigma$ deviation reported for the $2 \, ^3S$ ionization energy.

The correlated B-spline basis function (C-BSBF) method provides an alternative approach to high-accuracy calculations of Rydberg states. It inherits the linear independence and completeness of B-spline basis sets, which allow an accurate description of the long-range radial distribution of the Rydberg electron. In addition, explicit electron--electron correlation factors are incorporated into the basis functions to capture the two-electron correlation structure~\cite{YangSanJiang_2017, YangSanJiang_2019, Fanghao_2023, Fanghao_2024}. Recently, the method was extended to compute nonrelativistic energies of the $1snp \, ^{1,3}P$ states up to $n=27$ with at least 14 significant digits~\cite{Chijing_2025}. It was further applied to $1snp \, ^3P_J$ fine-structure intervals for $n=24$--37 with kHz-level accuracy, with additional results for $n=45$--51 obtained using extrapolation and fitting, enabling direct comparison with experiment and showing agreement with the available measurements~\cite{Fanghao_2026}. This provides a basis for extending the method to \emph{ab initio} calculations of ionization energies for helium Rydberg states.

In this work, the C-BSBF method is extended to calculate $^{1,3} P$ ionization energies of Rydberg states in $^4 \mathrm{He}$. High-precision \emph{ab initio} calculations are carried out for the states with $n=24$--35, treating the relativistic, recoil, QED, and higher-order corrections within a unified basis-set framework. The resulting ionization energies are used to determine quantum-defect parameters and to extrapolate to higher principal quantum numbers. Combined with the high-precision experimental $2S \rightarrow nP$ transition frequencies~\cite{Clausen_2021, Clausen_4He_2025}, the ionization energies of the metastable $2 \, ^1 S$ and $2 \, ^3 S$ states are obtained. The inverse fine-structure constant, the electron mass, the nuclear mass of $^4 \mathrm{He}$, and the Rydberg frequency are taken as $\alpha^{-1} = 137.035999177(21)$, $m_e = 1$, $M_0 = 7294.29954171(17) m_e$, and $c R_{\infty} = 3.2898419602500(36) \times 10^9$~MHz~\cite{CODATA_2022}, respectively.

\section{Theoretical Method}   

According to the theory of nonrelativistic quantum electrodynamics (NRQED)~\cite{Caswell_1986, Pachucki_2005, Pachucki_2006_energy}, the expansion of energy levels in powers of the fine structure constant $\alpha$ has the following form
\begin{equation}
    E(\alpha) = E^{(2)} + E^{(4)} + E^{(5)} + E^{(6)} + \cdots ,
\end{equation}
where $E^{(n)}$ denotes a contribution of order $m \alpha^n$ and may include powers of $\ln \alpha$. $E^{(2)} \equiv E_{\mathrm{nr}}$ is the nonrelativistic energy, $E^{(4)} \equiv E_{\mathrm{rel}}$ represents the leading relativistic correction, $E^{(5)} \equiv E_{\mathrm{QED}}$ corresponds to the leading QED correction, and $E^{(6)}$ contains the higher-order relativistic and QED contributions at order $m \alpha^6$.

\subsection{C-BSBFs}

The effective Hamiltonian for helium, derived after separation of the center-of-mass motion, can be expressed as
\begin{equation}
    H=\sum_{i=1}^2 \left( \frac{\mathbf{p}_i^2}{2 \mu}
    - \frac{Z}{r_i} \right)
    +\frac{1}{r_{12}}
    +\frac{\mathbf{p}_1 \cdot \mathbf{p}_2}{M_0} ,
\end{equation}
where $\mathbf{p}_i = - i \nabla_i$ is the momentum operator, $r_i$ is the coordinate of the $i$-th electron to the nucleus, $r_{12}=|\mathbf r_1-\mathbf r_2|$ is the interelectronic coordinate, $M_0$ is the nuclear mass, $\mu = m_e M_0/(m_e+M_0)$ is the reduced electron mass, and the nuclear charge $Z = 2$. The term $H_{\mathrm{rec}} = \mathbf{p}_1 \cdot \mathbf{p}_2/M_0$ represents the mass-polarization contribution due to nuclear motion in the center-of-mass frame, and the associated finite-nuclear-mass effects are extracted by comparison with the infinite-mass limit $M_0 \to \infty$.

For a given atomic state $n \, ^{2S+1} L$, characterized by the total spin angular momentum $S$, the total orbital angular momentum $L$, and magnetic quantum number $M$, the two-electron wave function of helium is expanded in C-BSBFs~\cite{Fanghao_2023, Fanghao_2024},
\begin{equation}
    \phi_{ij,c,\ell_1\ell_2}
    =\mathcal{A}\!\left[
    r_{12}^{c}\,
    B_{i}^{k}(r_{1})\,B_{j}^{k}(r_{2})\,
    \mathcal{Y}_{\ell_{1}\ell_{2}}^{LM}
    (\hat{\mathbf r}_{1},\hat{\mathbf r}_{2})
    \right],
\end{equation}
where $\mathcal{A}$ is the antisymmetrization operator for the two electrons, $c=0$ and $1$, and $B_i^{k}(r)$ ($i = 1, 2, \dots , N$) denotes the $i$-th B-spline of order $k$, where $N$ is the total number of B-splines confined within a cavity of radius $R_0$~\cite{bachau_2001}. $\mathcal{Y}_{\ell_{1}\ell_{2}}^{LM}(\hat{\mathbf r}_{1},\hat{\mathbf r}_{2})$ are the coupled spherical harmonics corresponding to orbital angular momenta $\ell_1$ and $\ell_2$~\cite{Brink_1994,Drake_1978}, subject to the restriction $\ell_{1},\ell_{2} \le \ell_{\max}$.

\subsection{Leading relativistic and QED corrections}

The leading relativistic correction of order $m\alpha^4$ to the nonrelativistic energy is given by the expectation value of the Breit-Pauli interaction~\cite{Drake_1992, Yerokhin_2010}:
\begin{equation}
    E_{\mathrm{rel}} = \left< H_{\mathrm{BP}} \right> ,
\end{equation}
$H_{\mathrm{BP}}$ can be decomposed into terms that do and do not give rise to the fine-structure splittings. The latter, including the reduced-mass dependence, is given by~\cite{Drake_1992}
\begin{equation}
    H_{\mathrm{nfs}}
    = \left( \frac{\mu}{m_e} \right)^3 \left( \frac{\mu}{m_e} H_1 + H_2 + H_4 + H_5 + \frac{m_e}{M_0} \Delta_2 \right) .  \label{eq:spin_independent_H}
\end{equation}
Here,
\begin{align}
    H_1      & = -\frac{\alpha^2}{8} \left( \mathbf{p}^4_1 + \mathbf{p}^4_2 \right), \nonumber                                      \\
    H_2      & = -\frac{\alpha^2}{2} \left[ \frac{1}{r_{12}} \mathbf{p}_1 \cdot \mathbf{p}_2
    + \frac{1}{r^3_{12}} \mathbf{r}_{12} \cdot \left( \mathbf{r}_{12} \cdot \mathbf{p}_1 \right) \mathbf{p}_2 \right] , \nonumber   \\
    H_4      & = \alpha^2 \pi \left[ \frac{Z}{2} \left( \delta^3(r_1) + \delta^3(r_2) \right) - \delta^3(r_{12}) \right], \nonumber \\
    H_5      & = -\alpha^2 \frac{8\pi}{3} \mathbf{s}_1 \cdot \mathbf{s}_2 \delta^3(r_{12}) \nonumber                                \\
    \Delta_2 & = -\frac{Z \alpha^2}{2} \sum_{i=1}^2 \left[ \frac{1}{r_i} \mathbf{P} \cdot \mathbf{p}_i
        + \frac{1}{r^3_i} \mathbf{r}_i \cdot \left( \mathbf{r}_i \cdot \mathbf{P} \right) \mathbf{p}_i \right] ,
\end{align}
where $\mathbf{P}=\mathbf{p}_1+\mathbf{p}_2$. $H_1$ originates from the relativistic correction to the electron kinetic energy, $H_2$ represents the orbit-orbit interaction, $H_4$ is the Darwin term that corresponds to the electron-nucleus and electron-electron contact interactions, $H_5$ describes the spin-spin contact interaction, and $\Delta_2$ arises from the transformation of the orbit-orbit interaction into relative coordinates.

The relativistic recoil correction associated with the reduced-mass dependence in Eq.~\eqref{eq:spin_independent_H} is~\cite{Pachucki_2015}
\begin{align}
    E_{M , 0}
    = \left< H_{\mathrm{nfs}} - H_{\mathrm{nfs},\infty} \right> ,
\end{align}
where $H_{\mathrm{nfs},\infty}$ denotes the part of the Breit-Pauli Hamiltonian that does not contribute to the fine-structure splittings, in the infinite nuclear-mass limit (with $\mu \to m_e$). The first-order recoil contribution $E_{M , 1}$ arises from the recoil addition to the Breit Hamiltonian and from the second-order perturbation correction involving $H_{\mathrm{nfs},\infty}$ and $H_{\mathrm{rec}}$~\cite{Pachucki_2015},
{\small
        \begin{align}
            E_{M , 1}
            = \left( \frac{\mu}{m_e} \right)^3 \frac{m_e}{M_0} \left< \Delta_2 \right> + 2 \frac{\mu}{m_e} \left< H_{\mathrm{nfs},\infty} \frac{1}{\left( E_0 - H \right)^{\prime}} H_{\mathrm{rec}} \right> . \label{eq:first_order_recoil}
        \end{align}
    }

The part of $H_{\mathrm{BP}}$ responsible for the fine-structure splittings is denoted by $H_{\mathrm{fs}}$. Its explicit form and the details of the fine-structure calculations have been presented in our previous work~\cite{Fanghao_2026} and will not be repeated here.

The leading QED correction can be written as an expectation value of the following effective operators~\cite{Araki_1957, Sucher_1958, Yerokhin_2010}:
\begin{align}
    E_{\mathrm{QED}}
     & = \alpha^3 \bigg\lbrace \frac{4Z}{3} \left[ \frac{19}{30} +  \ln \left( Z \alpha \right)^{-2} - \ln \frac{k_0}{Z^2} \right]  \nonumber                                      \\
     & \quad \times \langle \delta^3(r_1) + \delta^3(r_2) \rangle + \left[ \frac{164}{15} + \frac{14}{3} \ln \alpha \right] \nonumber                                              \\
     & \quad \times \langle \delta^3(r_{12}) \rangle - \frac{7}{6\pi} \langle r^{-3}_{12} \rangle + \langle H^{(5)}_{\mathrm{fs}} \rangle \bigg\rbrace , \label{eq:QED_correction}
\end{align}
where $\ln \left( k_0/Z^2 \right)$ is the Bethe logarithm (BL)~\cite{Schwartz_1961, Drake_2000}, $H^{(5)}_{\mathrm{fs}}$ represents the $m\alpha^5$ contribution to the fine-structure splittings~\cite{Yerokhin_2010, Fanghao_2026}, and $\langle r^{-3}_{12} \rangle$ denotes the Araki-Sucher correction~\cite{Korobov_2002, Yerokhin_2010}, defined as
\begin{equation}
    \langle r^{-3}_{12} \rangle = \lim_{a \to 0} \langle r^{-3}_{12} \Theta(r_{12}-a)
    + 4\pi \left( \gamma + \ln a \right) \delta^3(r_{12}) \rangle,
\end{equation}
where $\Theta(x)$ and $\gamma$ are the step function and the Euler constant, respectively. Owing to its singular nature, the Araki-Sucher term is transformed into an equivalent but more regular form using the global operator method proposed by Drachman~\cite{Drachman_1981}, which takes the following form~\cite{Pachucki_2004, Fanghao_2023}:
\begin{align}
    \langle r^{-3}_{12} \rangle & = - \sum_{i=1}^{2} \langle \nabla_i \psi \left| r^{-1}_{12} \ln r_{12} \right| \nabla_i \psi \rangle \nonumber                                                \\
                                & \quad + \langle \psi | 2 ( E_\psi - V ) \frac{\ln r_{12}}{r_{12}} + 4 \pi \left( 1 + \gamma \right) \delta^3(r_{12}) | \psi \rangle , \label{eq:Araki_Sucher}
\end{align}
where $E_\psi$ is the corresponding eigenvalue of the two-electron wave function $\psi$, and $V=-Z/r_1-Z/r_2+1/r_{12}$ is the nonrelativistic Coulomb potential.

\subsection{$m \alpha^6$ correction}

The $m\alpha^6$ correction generally consists of several components, including the logarithmic term, the second-order perturbation contribution induced by the Breit-Pauli Hamiltonian, the nonradiative effective Hamiltonian, the one- and two-loop radiative terms, and the spin-dependent contributions~\cite{Pachucki_2006_energy, Yerokhin_2010}. The dominant contribution comes from the second-order singlet-triplet mixing between the $n \, ^3P_1$ and $n \, ^1P_1$ states. This effect arises from the spin-dependent Breit-Pauli interaction. The singlet-triplet contribution $E_{\mathrm{st}}$ is obtained by exact diagonalization of an effective Hamiltonian~\cite{Drake_1979, Drake_1992, Drake_1988}. In the present calculation, an effective $6 \times 6$ Hamiltonian is constructed in a six-state basis. Taking the $n \, ^3 P$ state as an example, this basis consists of the two isoconfigurational states, $| n \, ^3 P \rangle$ and $| n \, ^1 P \rangle$, together with the four nearest non-isoconfigurational states of opposite spin symmetry, $|(n+1) \, ^1 P \rangle$, $| (n+2) \, ^1 P \rangle$, $|(n-1) \, ^1 P \rangle$ and $| (n-2) \, ^1 P \rangle$. The details of the calculation can be found in Ref.~\cite{Fanghao_2026}.
%

In addition, the partial radiative corrections corresponding to the one- and two-loop electron self-energy and vacuum-polarization terms, $E_{\mathrm{R1}}$ and $E_{\mathrm{R2}}$, are included~\cite{Yelkhovsky_2001, Pachucki_2006, Pachucki_2006_energy}
\begin{align}
    E_{\mathrm{R1}}
     & = \alpha^4 \bigg\lbrace Z^2 \left[ \frac{427}{96} - 2 \ln(2) \right] \pi \langle \delta^3(r_1) + \delta^3(r_2) \rangle \nonumber                                      \\
     & \quad + \left[ \frac{6 \zeta(3)}{\pi^2} - \frac{697}{27 \pi^2} - 8\ln(2) + \frac{1099}{72} \right] \pi \langle \delta^3(r_{12}) \rangle \bigg\rbrace , \label{eq:ER1}
\end{align}
\begin{align}
    E_{\mathrm{R2}}
     & = \alpha^4 \bigg\lbrace Z \left[ -\frac{9\zeta(3)}{4\pi^2} - \frac{2179}{648\pi^2} + \frac{3 \ln(2)}{2} - \frac{10}{27} \right] \nonumber \\
     & \quad \times \pi \langle \delta^3(r_1) + \delta^3(r_2) \rangle + \bigg[ \frac{15\zeta(3)}{2\pi^2} + \frac{631}{54\pi^2} \nonumber         \\
     & \quad  - 5\ln(2) + \frac{29}{27} \bigg] \pi \langle \delta^3(r_{12}) \rangle \bigg\rbrace , \label{eq:ER2}
\end{align}
where $\zeta$ is the Riemann zeta function. The remaining contributions of order $m\alpha^6$, including the logarithmic term, other second-order perturbative contributions, and the expectation value of the nonradiative effective Hamiltonian, are only estimated in the present work.


\section{Results and Discussion}

\subsection{Nonrelativistic energies}

In our previous work~\cite{Fanghao_2026}, a cavity radius of $R_0=4000$ a.u. was employed to calculate the fine-structure splittings for states up to $n=37$. Using the same set of parameters, we extend the calculations to the ionization energies of Rydberg states, covering several experimentally measured levels ($n \, ^1 P$: $n=24$--35, $n \, ^3 P$: $n=27$--35). Calculations were performed using B-splines of sizes $N = 80$, 90, and 100 at $\ell_{\max} = 4$, supplemented by an additional calculation with $N = 90$ and $\ell_{\max} = 5$ to evaluate the influence of the $\ell_{\max}$ truncation. The largest calculation corresponds to a Hamiltonian matrix dimension of \num{81000}. To ensure numerical accuracy and computational efficiency, a self-developed quadruple (double-double) precision parallel program was employed, with the largest basis set requiring about 20 hours using 128 cores in parallel. The uncertainty is estimated following Ref.~\cite{Fanghao_2026} as the larger of the two sources of uncertainty: (i) the maximum deviation between the extrapolated value and the results obtained with the three largest basis sets ($N = 80$, 90, and 100 at $\ell_{\max} = 4$), and (ii) the change in the result upon increasing $\ell_{\max}$ from 4 to 5.

The nonrelativistic energies of $n \, ^{1,3}P$ states ($n=24$--35) of $^{\infty} \mathrm{He}$ are presented in Table~\ref{tab:nonrel-energy}. The calculated results have converged to at least 14 significant digits. For comparison, high-accuracy values obtained using the Hylleraas basis~\cite{Drake_2026} are also listed, and the present results agree within the uncertainties. For lower principal quantum numbers ($n=24$--27), the agreement is found to reach up to 18 significant digits, indicating that the C-BSBF method achieves very high numerical accuracy. As $n$ increases, the accuracy of the C-BSBF results slightly degrades due to the limitations imposed by the finite cavity and the spline basis size, while still showing agreement with the Hylleraas results to about 14 digits. This accuracy level is adequate to enable subsequent calculations of relativistic and QED corrections, and the agreement between the C-BSBF and Hylleraas results validates the reliability of the present numerical method for describing Rydberg states.

\setlength{\cmidrulewidth}{0.4pt}
\begin{table*}[ht]
    \scriptsize
    \caption{\label{tab:nonrel-energy} Nonrelativistic energies of helium $n \,^{1,3} P$ states of $^{\infty} \mathrm{He}$ (in a.u.). The calculations adopt knot parameter $\tau = 0.00228$, spline order $k=15$, $\ell_{\max}=4$, and cavity radius $R_0=4000\,\mathrm{a.u.}$. The numbers in parentheses are the uncertainties.}
    \begin{threeparttable}
        \begin{tabular*}{\textwidth}{@{\extracolsep{\fill}} c d{4.24} d{2.30} d{4.24} d{2.30}}
            \toprule
            \multirow{2}{*}{$n$} & \multicolumn{2}{c}{$^1 P$} & \multicolumn{2}{c}{$^3 P$}    \\
            \cmidrule(lr){2-3} \cmidrule(lr){4-5}
            & \multicolumn{1}{c}{C-BSBF} & \multicolumn{1}{c}{Hylleraas~\cite{Drake_2026}}    & \multicolumn{1}{c}{C-BSBF}   & \multicolumn{1}{c}{Hylleraas~\cite{Drake_2026}}           \\
            \midrule
            24 & -2.000\,867\,180\,846\,170\,11(1)     & -2.000\,867\,180\,846\,170\,111\,282\,23(6)    & -2.000\,873\,014\,566\,616\,659(1)     & -2.000\,873\,014\,566\,616\,659\,392\,40(9)       \\
            25 & -2.000\,799\,226\,024\,103\,06(1)     & -2.000\,799\,226\,024\,103\,063\,045\,555(6)   & -2.000\,804\,386\,829\,929\,070(6)     & -2.000\,804\,386\,829\,929\,070\,608\,458(13)     \\
            26 & -2.000\,738\,956\,837\,741\,71(2)     & -2.000\,738\,956\,837\,741\,719\,217\,65(4)    & -2.000\,743\,544\,360\,330\,30(2)      & -2.000\,743\,544\,360\,330\,295\,812\,81(17)      \\
            27 & -2.000\,685\,256\,528\,882\,40(5)     & -2.000\,685\,256\,528\,882\,402\,121\,453(18)  & -2.000\,689\,352\,623\,570\,98(4)      & -2.000\,689\,352\,623\,570\,976\,558\,76(3)       \\
            28 & -2.000\,637\,204\,047\,170\,8(1)      & -2.000\,637\,204\,047\,170\,837\,143\,041(29)  & -2.000\,640\,876\,467\,041\,02(8)      & -2.000\,640\,876\,467\,041\,024\,093\,026(20)     \\
            29 & -2.000\,594\,034\,290\,981\,6(3)      & -2.000\,594\,034\,290\,981\,601\,789\,1(1)     & -2.000\,597\,339\,504\,545\,6(2)       & -2.000\,597\,339\,504\,545\,631\,854\,5(2)        \\
            30 & -2.000\,555\,107\,462\,373\,0(8)      & -2.000\,555\,107\,462\,372\,974\,259\,1(24)    & -2.000\,558\,092\,835\,758\,0(8)       & -2.000\,558\,092\,835\,757\,975\,283\,2(5)        \\
            31 & -2.000\,519\,885\,224\,315(3)         & -2.000\,519\,885\,224\,314\,856\,988\,9(1)     & -2.000\,522\,590\,726\,605(2)          & -2.000\,522\,590\,726\,604\,946\,168\,6(4)        \\
            32 & -2.000\,487\,911\,987\,757(8)         & -2.000\,487\,911\,987\,756\,799\,324\,1(1)     & -2.000\,490\,371\,534\,948(7)          & -2.000\,490\,371\,534\,947\,756\,033\,9(3)        \\
            33 & -2.000\,458\,800\,104\,87(2)          & -2.000\,458\,800\,104\,867\,467\,859\,0(2)     & -2.000\,461\,042\,627\,37(2)           & -2.000\,461\,042\,627\,369\,818\,585\,4(4)        \\
            34 & -2.000\,432\,218\,063\,63(7)          & -2.000\,432\,218\,063\,626\,956\,609\,8(4)     & -2.000\,434\,268\,360\,44(7)           & -2.000\,434\,268\,360\,440\,676\,552\,3(3)        \\
            35 & -2.000\,407\,881\,008\,1(2)           & -2.000\,407\,881\,008\,092\,820\,968(2)        & -2.000\,409\,760\,435\,0(2)            & -2.000\,409\,760\,435\,014\,704\,045(2)           \\
            \bottomrule
        \end{tabular*}
    \end{threeparttable}
\end{table*}

\subsection{Leading relativistic correction}

Based on the nonrelativistic energies and wave functions, the expectation values of the operators entering the relativistic and recoil corrections are computed and used to evaluate the corresponding operator contributions with finite nuclear mass. The quantity $B^{\prime}_M$ denotes the finite nuclear-mass expectation value, given by~\cite{Pachucki_2015}
\begin{align}
    B^\prime_M = \mu^m B_{\infty} + \frac{\mu}{M_0} B_M,
\end{align}
where $m=4$ for $p^4_1$ and $m=3$ for all other operators. The quantity $B_{\infty}$ denotes the expectation value for infinite nuclear mass, while $B_M$ represents the mass-dependent correction arising from the second-order perturbation contribution to the recoil correction in Eq.~\eqref{eq:first_order_recoil}, induced by the mass-polarization operator~\cite{Pachucki_2015}
\begin{align}
    B_M = 2 \left< B_{\infty} \frac{1}{\left( E_0 - H \right)^{\prime}} \left( \mathbf{p}_1 \cdot \mathbf{p}_2 \right) \right> .
\end{align}

The numerical values of $B_{\infty}$ and $B_M$ for the operators considered are listed in Tables~\ref{tab:malpha4_operator_1P} and~\ref{tab:malpha4_operator_3P}. The infinite-mass results $B_{\infty}$ are accurate to at least 7 significant digits, while the finite-mass results $B_M$ are converged to at least 4 significant digits. Taking the $24 \, ^1 P$ state as an example, $p^4_1$ at infinite nuclear mass is \num{40.0001218312(1)}, which is in agreement with the high-precision value \num{40.00012183120973(22)} reported by Drake \emph{et al.}~\cite{Drake_2026}. Upon inclusion of the mass-dependent correction, one obtains $p^{4 \prime}_1=$ \num{39.9781944171(1)}, consistent with the value \num{39.97819441727845(58)} reported by Drake \emph{et al.} ~\cite{Drake_2026}. Given that the relativistic and recoil corrections are determined by these operator expectation values, the achieved accuracy is sufficient to ensure kHz precision for both corrections.

\begin{table*}[ht]
    \footnotesize
    \caption{\label{tab:malpha4_operator_1P} Expectation values of the operators $p^4_1$, $\delta^3(r_1)$, $\delta(r_{12})$, $H_2$, and $\Delta_2$ (first three in a.u., last two in $\alpha^2$ a.u.) for $n \, ^1 P$ states of $\mathrm{He}$. For each principal quantum number, the first and second lines correspond to $B_{\infty}$ and $B_M$, respectively. The numbers in parentheses and square brackets are the uncertainties and powers of 10, respectively.}
    \begin{threeparttable}
        \begin{tabular*}{\textwidth}{@{\extracolsep{\fill}} c d{4.16} d{4.22} d{4.15} d{4.18}  d{4.22}}
            \toprule
            \multicolumn{1}{c}{$n$} & \multicolumn{1}{c}{$\langle p^4_1 \rangle$} & \multicolumn{1}{c}{$\langle \delta^3(r_1) \rangle$} & \multicolumn{1}{c}{$\langle \delta^3(r_{12}) \rangle$} & \multicolumn{1}{c}{$\langle H_2 \rangle$} & \multicolumn{1}{c}{$\langle \Delta_2 \rangle$}          \\
            \midrule
            24 & 40.000\,121\,831\,2(1)           &  1.273\,240\,319\,814\,349(6)     &  5.480\,662(1)[-7]       & -1.407\,587\,0(1)[-5]     & -16.000\,247\,772\,220\,027\,7(3)    \\
            \, & 5.383\,3(1)[-4]                  & 2.145\,505\,701\,9(3)[-5]         & -1.987(1)[-6]            & 8.399\,715\,1(1)[-5]      &                                      \\
            25 & 40.000\,107\,954\,5(1)           &  1.273\,240\,230\,502\,467(5)     &  4.849\,793(1)[-7]       & -1.245\,507\,7(1)[-5]     & -16.000\,219\,444\,162\,306\,1(2)    \\
            \, & 4.764\,6(1)[-4]                  & 1.898\,331\,624\,8(5)[-5]         & -1.757(1)[-6]            & 7.439\,768\,9(1)[-5]      &                                      \\
            26 & 40.000\,096\,107\,5(1)           &  1.273\,240\,154\,402\,268(4)     &  4.312\,131(1)[-7]       & -1.107\,384\,2(1)[-5]     & -16.000\,195\,274\,619\,216\,3(2)    \\
            \, & 4.237\,2(1)[-4]                  & 1.687\,717\,078\,6(2)[-5]         & -1.562(1)[-6]            & 6.620\,698\,0(1)[-5]      &                                      \\
            27 & 40.000\,085\,932\,2(1)           &  1.273\,240\,089\,158\,32(1)      &  3.851\,083(1)[-7]       & -9.889\,496(1)[-6]        & -16.000\,174\,527\,743\,150\,5(2)    \\
            \, & 3.784\,9(1)[-4]                  & 1.507\,144\,376\,3(3)[-5]         & -1.394(1)[-6]            & 5.917\,579\,8(1)[-5]      &                                      \\
            28 & 40.000\,077\,143\,9(1)           &  1.273\,240\,032\,901\,733(3)     &  3.453\,475(1)[-7]       & -8.868\,175(1)[-6]        & -16.000\,156\,618\,553\,471\,8(5)    \\
            \, & 3.394\,8(1)[-4]                  & 1.351\,442\,030\,6(2)[-5]         & -1.249(1)[-6]            & 5.310\,600\,6(1)[-5]      &                                      \\
            29 & 40.000\,069\,514\,2(1)           &  1.273\,239\,984\,136\,724(2)     &  3.108\,764(1)[-7]       & -7.982\,769(1)[-6]        & -16.000\,141\,078\,051\,275(1)       \\
            \, & 3.056\,5(1)[-4]                  & 1.216\,471\,188\,2(4)[-5]         & -1.124(1)[-6]            & 4.783\,874\,9(1)[-5]      &                                      \\
            30 & 40.000\,062\,858\,2(1)           &  1.273\,239\,941\,656\,606(2)     &  2.808\,438(1)[-7]       & -7.211\,403(1)[-6]        & -16.000\,127\,527\,256\,107(3)       \\
            \, & 2.761\,7(1)[-4]                  & 1.098\,892\,832\,4(5)[-5]         & -1.015(1)[-6]            & 4.324\,565\,6(1)[-5]      &                                      \\
            31 & 40.000\,057\,025\,5(1)           &  1.273\,239\,904\,480\,619(2)     &  2.545\,579(1)[-7]       & -6.536\,294(1)[-6]        & -16.000\,115\,657\,677\,72(1)        \\
            \, & 2.503\,6(1)[-4]                  & 9.959\,930\,99(1)[-6]             & -9.196(1)[-7]            & 3.922\,222\,7(1)[-5]      &                                      \\
            32 & 40.000\,051\,892\,7(1)           &  1.273\,239\,871\,806\,073(6)     &  2.314\,521(1)[-7]       & -5.942\,885(1)[-6]        & -16.000\,105\,216\,482\,86(3)        \\
            \, & 2.276\,7(1)[-4]                  & 9.055\,508\,60(3)[-6]             & -8.358(1)[-7]            & 3.568\,281\,5(1)[-5]      &                                      \\
            33 & 40.000\,047\,357\,7(1)           &  1.273\,239\,842\,971\,75(1)      &  2.110\,599(1)[-7]       & -5.419\,181(1)[-6]        & -16.000\,095\,995\,124\,45(9)        \\
            \, & 2.076\,4(1)[-4]                  & 8.257\,364\,29(7)[-6]             & -7.619(1)[-7]            & 3.255\,677\,2(1)[-5]      &                                      \\
            34 & 40.000\,043\,336\,1(1)           &  1.273\,239\,817\,429\,64(3)      &  1.929\,943(1)[-7]       & -4.955\,244(1)[-6]        & -16.000\,087\,820\,550\,0(3)         \\
            \, & 1.899\,0(1)[-4]                  & 7.550\,333\,8(2)[-6]              & -6.965(1)[-7]            & 2.978\,547\,0(1)[-5]      &                                      \\
            35 & 40.000\,039\,757\,3(1)           &  1.273\,239\,794\,722\,96(6)      &  1.769\,328(1)[-7]       & -4.542\,784(1)[-6]        & -16.000\,080\,548\,349\,3(9)         \\
            \, & 1.741\,2(1)[-4]                  & 6.921\,776\,9(2)[-6]              & -6.383(1)[-7]            & 2.731\,998\,3(1)[-5]      &                                      \\
            \bottomrule
        \end{tabular*}
    \end{threeparttable}
\end{table*}

\begin{table*}[ht]
    \footnotesize
    \caption{\label{tab:malpha4_operator_3P} Expectation values of the operators $p^4_1$, $\delta^3(r_1)$, $H_2$, and $\Delta_2$ (first three in a.u., last two in $\alpha^2$ a.u.) for $n \, ^3 P$ states of $\mathrm{He}$. For each principal quantum number, the first and second lines correspond to $B_{\infty}$ and $B_M$, respectively. The numbers in parentheses and square brackets are the uncertainties and powers of 10, respectively.}
    \begin{threeparttable}
        \begin{tabular*}{\textwidth}{@{\extracolsep{\fill}} c d{4.16} d{4.22} d{4.18}  d{4.22}}
            \toprule
            \multicolumn{1}{c}{$n$} & \multicolumn{1}{c}{$\langle p^4_1 \rangle$} & \multicolumn{1}{c}{$\langle \delta^3(r_1) \rangle$} & \multicolumn{1}{c}{$\langle H_2 \rangle$} & \multicolumn{1}{c}{$\langle \Delta_2 \rangle$}          \\
            \midrule
            24 & 39.999\,877\,247\,827(1)         & 1.273\,232\,049\,261\,252(2)         & 1.915\,389\,343(1)[-5]  & -15.999\,943\,996\,684\,330\,88(7)      \\
            \, & -3.427\,13(1)[-4]                & -1.719\,307\,523\,0(1)[-5]           & 8.511\,110\,692(1)[-5]  &                                         \\
            25 & 39.999\,891\,603\,099(1)         & 1.273\,232\,915\,607\,791(1)         & 1.694\,034\,035(1)[-5]  & -15.999\,950\,688\,424\,592\,01(3)      \\
            \, & -3.026\,321(1)[-4]               & -1.518\,441\,066\,2(4)[-5]           & 7.533\,716\,744(1)[-5]  &                                         \\
            26 & 39.999\,903\,804\,540(1)         & 1.273\,233\,653\,420\,354(1)         & 1.505\,514\,385(1)[-5]  & -15.999\,956\,356\,055\,648\,92(6)      \\
            \, & -2.685\,726(1)[-4]               & -1.347\,702\,630\,4(1)[-5]           & 6.700\,446\,950(1)[-5]  &                                         \\
            27 & 39.999\,914\,241\,395(1)         & 1.273\,234\,285\,679\,284(1)         & 1.343\,960\,853(1)[-5]  & -15.999\,961\,188\,080\,536\,89(4)      \\
            \, & -2.394\,434(1)[-4]               & -1.201\,645\,932\,8(4)[-5]           & 5.985\,682\,721(1)[-5]  &                                         \\
            28 & 39.999\,923\,221\,461(1)         & 1.273\,234\,830\,605\,630(1)         & 1.204\,719\,247(1)[-5]  & -15.999\,965\,332\,880\,580\,8(2)       \\
            \, & -2.143\,830(1)[-4]               & -1.075\,965\,238\,29(8)[-5]          & 5.369\,082\,022(1)[-5]  &                                         \\
            29 & 39.999\,930\,990\,289(1)         & 1.273\,235\,302\,772\,535\,1(6)      & 1.084\,066\,982(1)[-5]  & -15.999\,968\,908\,329\,226\,8(2)       \\
            \, & -1.927\,044(1)[-4]               & -9.672\,260\,285(8)[-6]              & 4.834\,353\,618(1)[-5]  &                                         \\
            30 & 39.999\,937\,745\,397(1)         & 1.273\,235\,713\,929\,644\,2(9)      & 9.790\,025\,832(1)[-6]  & -15.999\,972\,008\,862\,144\,7(7)       \\
            \, & -1.738\,554(1)[-4]               & -8.726\,660\,011(3)[-6]              & 4.368\,346\,263(1)[-5]  &                                         \\
            31 & 39.999\,943\,646\,875(1)         & 1.273\,236\,073\,621\,265\,8(6)      & 8.870\,878\,401(1)[-6]  & -15.999\,974\,710\,736\,054(2)          \\
            \, & -1.573\,886(1)[-4]               & -7.900\,463\,87(3)[-6]               & 3.960\,364\,649(1)[-5]  &                                         \\
            32 & 39.999\,948\,825\,412(1)         & 1.273\,236\,389\,654\,578(2)         & 8.063\,282\,454(1)[-6]  & -15.999\,977\,075\,981\,785(7)          \\
            \, & -1.429\,390(1)[-4]               & -7.175\,397\,10(3)[-6]               & 3.601\,650\,511(1)[-5]  &                                         \\
            33 & 39.999\,953\,388\,402(1)         & 1.273\,236\,668\,457\,589(3)         & 7.350\,815\,785(1)[-6]  & -15.999\,979\,155\,405\,14(2)           \\
            \, & -1.302\,067(1)[-4]               & -6.536\,442\,3(1)[-6]                & 3.284\,985\,354(2)[-5]  &                                         \\
            34 & 39.999\,957\,424\,651(1)         & 1.273\,236\,915\,355\,21(1)          & 6.719\,873\,97(2)[-6]   & -15.999\,980\,990\,885\,87(6)           \\
            \, & -1.189\,438(1)[-4]               & -5.971\,181\,5(2)[-6]                & 3.004\,383\,636(7)[-5]  &                                         \\
            35 & 39.999\,961\,008\,023(1)         & 1.273\,237\,134\,783\,96(2)          & 6.159\,122\,04(6)[-6]   & -15.999\,982\,617\,154\,2(2)            \\
            \, & -1.089\,442(1)[-4]               & -5.469\,285\,9(5)[-6]                & 2.754\,853\,89(2)[-5]   &                                         \\
            \bottomrule
        \end{tabular*}
    \end{threeparttable}
\end{table*}

\subsection{Leading QED correction}

In addition to the operators involved in the relativistic contributions discussed above, the evaluation of QED corrections requires two additional terms corresponding to the Araki-Sucher and BL contributions. The corresponding numerical results are listed in Table~\ref{tab:malpha5_operator}. $\left< r^{-3}_{12} \right>$ is highly sensitive to short-range electron-electron correlations. For Rydberg states, its magnitude is on the order of $10^{-5}$, indicating that the probability of finding the two electrons at short interelectronic distances is strongly suppressed as the Rydberg electron becomes more spatially extended. Owing to its singular nature, this term requires an accurate description of the short-range behavior of the wave function. By evaluating it with Eq.~\eqref{eq:Araki_Sucher}, the singularity is effectively reduced, and an accuracy of about 7 significant digits is achieved, which is sufficient for QED corrections.

Direct \emph{ab initio} calculation of the BL becomes increasingly difficult for Rydberg states. The spatial extent of the Rydberg electron increases rapidly with $n$. This imposes more stringent requirements on the completeness of the intermediate-state expansion over a large spatial range. The asymptotic expansion form~\cite{Drake_2001, Korobov_2019} is therefore adopted in the present work. This approach takes advantage of the high-$n$ asymptotic behavior of the BL, substantially reducing the numerical complexity while preserving sufficient accuracy. Using the fitting parameters of Ref.~\cite{Drake_2001}, the calculated values are also listed in Table~\ref{tab:malpha5_operator}. Since only the nonrecoil part of the QED corrections is considered in the present work, finite-mass contributions are not included in the fitted evaluation of the BL.

\setlength{\cmidrulewidth}{0.4pt}
\begin{table*}[ht]
    \footnotesize
    \caption{\label{tab:malpha5_operator} Expectation values of $r^{-3}_{12}$ and Bethe logarithm used for $n \, ^{1,3} P$ states of $^{\infty}\mathrm{He}$. The numbers in parentheses and square brackets are the uncertainties and powers of 10, respectively.}
    \begin{threeparttable}
        \begin{tabular*}{\textwidth}{@{\extracolsep{\fill}} c d{4.18} d{2.15} d{4.18} d{2.15}}
            \toprule
            \multirow{2}{*}{$n$} & \multicolumn{2}{c}{$^1 P$} & \multicolumn{2}{c}{$^3 P$}    \\
            \cmidrule(lr){2-3} \cmidrule(lr){4-5}
            & \multicolumn{1}{c}{$\langle 1/r^3_{12} \rangle$} & \multicolumn{1}{c}{BL}    & \multicolumn{1}{c}{$\langle 1/r^3_{12} \rangle$}   & \multicolumn{1}{c}{BL}           \\
            \midrule
            24 & 2.462\,504(1)[-5]   & 2.984\,128\,212\,4(4)         & 2.296\,066\,1(1)[-5]       & 2.984\,128\,078(1)         \\
            25 & 2.178\,790(1)[-5]   & 2.984\,128\,251\,5(3)         & 2.030\,605\,0(1)[-5]       & 2.984\,128\,132(1)         \\
            26 & 1.937\,037(1)[-5]   & 2.984\,128\,284\,9(3)         & 1.804\,541\,8(1)[-5]       & 2.984\,128\,179(1)         \\
            27 & 1.729\,768(1)[-5]   & 2.984\,128\,313\,6(3)         & 1.610\,830\,0(1)[-5]       & 2.984\,128\,219(1)         \\
            28 & 1.551\,045(1)[-5]   & 2.984\,128\,338\,4(2)         & 1.443\,883\,0(1)[-5]       & 2.984\,128\,253\,9(9)      \\
            29 & 1.396\,121(1)[-5]   & 2.984\,128\,359\,9(2)         & 1.299\,233\,0(1)[-5]       & 2.984\,128\,284\,0(8)      \\
            30 & 1.261\,161(1)[-5]   & 2.984\,128\,378\,7(2)         & 1.173\,278\,3(1)[-5]       & 2.984\,128\,310\,1(8)      \\
            31 & 1.143\,051(1)[-5]   & 2.984\,128\,395\,1(2)         & 1.063\,093\,4(1)[-5]       & 2.984\,128\,333\,1(7)      \\
            32 & 1.039\,239(1)[-5]   & 2.984\,128\,409\,6(2)         & 9.662\,854(1)[-6]          & 2.984\,128\,353\,2(6)      \\
            33 & 9.476\,28(1)[-6]    & 2.984\,128\,422\,4(1)         & 8.808\,843(1)[-6]          & 2.984\,128\,371\,0(6)      \\
            34 & 8.664\,75(1)[-6]    & 2.984\,128\,433\,7(1)         & 8.052\,581(1)[-6]          & 2.984\,128\,386\,8(5)      \\
            35 & 7.943\,31(1)[-6]    & 2.984\,128\,443\,8(1)         & 7.380\,473(1)[-6]          & 2.984\,128\,400\,8(5)      \\
            \bottomrule
        \end{tabular*}
    \end{threeparttable}
\end{table*}

\subsection{$m \alpha^6$ correction}

The dominant part of the $m \alpha^6$ correction comes from the singlet-triplet mixing. The corresponding results for the $n \, ^{1,3} P$ states ($n=24$--35) are listed in Table~\ref{tab:malpha6_ST}. For the $^3P$ states, only the $J=1$ fine-structure component contributes to the singlet-triplet mixing~\cite{Drake_1979, Drake_1988, Fanghao_2026}. For the centroid level, the corresponding contribution is obtained as the weighted average of the fine-structure components. Table~\ref{tab:malpha6_ST} gives the singlet-triplet mixing results obtained by including only the isoconfigurational contribution and by additionally including the four nearest opposite-spin levels. For the isoconfigurational contribution, the corresponding values for the $24 \, ^1 P$ and $27 \, ^1P$ states are \num{2.271769} and \num{1.594654}~kHz, respectively, in agreement with the previously reported values of \num{2.2762} and \num{1.60}~kHz~\cite{Bondy_2025, Drake_2026}. The effect of the four additional levels can be illustrated by the $24 \, ^3P$ state, for which they provide an additional contribution of $-0.028$~kHz. The results including both the isoconfigurational term and the four nearest neighboring levels are adopted in the subsequent calculations unless otherwise stated.

In addition to the singlet-triplet mixing effect, the one- and two-loop radiative terms $E_{\mathrm{R1}}$ and $E_{\mathrm{R2}}$ are also taken into account. With the required contact expectation values already listed in Tables~\ref{tab:malpha4_operator_1P} and \ref{tab:malpha4_operator_3P}, these two terms can be obtained directly from Eqs.~\eqref{eq:ER1} and \eqref{eq:ER2}, and their numerical values are therefore not listed separately here.

\begin{table}[ht]
    \footnotesize
    \caption{\label{tab:malpha6_ST} Singlet-triplet mixing correction for the $n \, ^{1,3} P$ states of $^{\infty}\mathrm{He}$ (in kHz). The notation ``Iso.'' denotes the isoconfigurational contribution, and ``Iso+Non-iso.'' denotes the combined contribution from the isoconfigurational term and four additional nearest opposite-spin levels. }
    \begin{threeparttable}
        \begin{tabular}{l d{4.9} d{3.9} d{4.8} d{4.8}}
            \toprule
            \multirow{2}{*}{$n$}  & \multicolumn{2}{c}{$^1 P$} & \multicolumn{2}{c}{$^3 P$}                                                                     \\
            \cmidrule(lr){2-3} \cmidrule(lr){4-5}
            \multicolumn{1}{c}{ } & \multicolumn{1}{c}{Iso.}   & \multicolumn{1}{c}{Iso+Non-iso.} & \multicolumn{1}{c}{Iso.} & \multicolumn{1}{c}{Iso+Non-iso.} \\
            \midrule
            24                    & 2.271\,769                 & 2.280\,971                       & -0.757\,256              & -0.729\,574                      \\
            25                    & 2.009\,511                 & 2.016\,001                       & -0.669\,837              & -0.645\,899                      \\
            26                    & 1.786\,120                 & 1.790\,536                       & -0.595\,373              & -0.574\,546                      \\
            27                    & 1.594\,654                 & 1.597\,480                       & -0.531\,551              & -0.513\,328                      \\
            28                    & 1.429\,607                 & 1.431\,211                       & -0.476\,536              & -0.460\,507                      \\
            29                    & 1.286\,574                 & 1.287\,241                       & -0.428\,858              & -0.414\,692                      \\
            30                    & 1.162\,005                 & 1.161\,952                       & -0.387\,335              & -0.374\,759                      \\
            31                    & 1.053\,013                 & 1.052\,411                       & -0.351\,004              & -0.339\,792                      \\
            32                    & 0.957\,237                 & 0.956\,217                       & -0.319\,079              & -0.309\,044                      \\
            33                    & 0.872\,734                 & 0.871\,400                       & -0.290\,911              & -0.281\,897                      \\
            34                    & 0.797\,892                 & 0.796\,324                       & -0.265\,964              & -0.257\,838                      \\
            35                    & 0.731\,369                 & 0.729\,632                       & -0.243\,790              & -0.236\,442                      \\
            \bottomrule
        \end{tabular}
    \end{threeparttable}
\end{table}

\subsection{Remainder contributions}

The remaining contributions of order $m \alpha^6$, together with the corrections of order $m \alpha^7$, are generally difficult to evaluate directly. For Rydberg states, these terms may be estimated following the procedure adopted in Refs.~\cite{Bondy_2025, Drake_2026}. Specifically, one first evaluates the dominant contributions at order $m \alpha^6$ that can be obtained relatively straightforwardly. For the $2 \, ^{1,3} P$ states, these directly evaluable terms are summarized in Table~\ref{tab:malpha6_remainder}. For $E_{\mathrm{st}}$, only the isoconfigurational singlet-triplet mixing is included, and the $m\alpha^6\ln(\alpha)$ term is taken from Ref.~\cite{Drake_1988}. Taking the $2 \, ^1 P$ state as an example, the sum of these directly evaluable contributions is \num{6.9537}~MHz, whereas the complete correction of order $m \alpha^6$ for the $2 \, ^1 P$ state is \num{8.818}~MHz~\cite{Pachucki_2017}. Including also the correction of order $m \alpha^7$, the combined $m \alpha^6$ and $m \alpha^7$ contribution becomes \num{9.628} MHz~\cite{Pachucki_2017, Pachucki_2021}. Subtracting the directly evaluable terms listed above from \num{9.628}~MHz leaves a remaining contribution of $E_{\mathrm{rmdr}}=2.674$~MHz. For higher Rydberg states, this remainder can be further assumed to scale approximately as $1/n^3$. As an illustration, the corresponding remainder for the $24 \, ^1 P$ state can be estimated as $2.674 \times \left( 2/24 \right)^3 = 1.547$~kHz. The scaled remainder $E_{\mathrm{rmdr}}$ for each Rydberg state is then included in the final results, with its full value taken as an uncertainty estimate.

\begin{table}[ht]
    \footnotesize
    \caption{\label{tab:malpha6_remainder} Evaluation of the dominant contributions and the remainder for the $2 \, ^{1,3} P$ states of $^{\infty} \mathrm{He}$ (in MHz). $E_{\mathrm{st}}$
    denotes the isoconfigurational singlet-triplet mixing contribution, $E_{\mathrm{R1}}$ and $E_{\mathrm{R2}}$ are the one- and two-loop radiative terms. The $m \alpha^6 \ln (\alpha)$ term is taken from Ref.~\cite{Drake_1993}, while $E^{(6)}$ and $E^{(7)}$ are adopted from Refs.~\cite{Pachucki_2017, Pachucki_2021}. The remainder is defined as $E_{\mathrm{rmdr}}=E^{(6)}+E^{(7)}-(E_{\mathrm{st}}+E_{\mathrm{R1}}+E_{\mathrm{R2}}+m \alpha^6 \ln (\alpha))$. The numbers in square brackets represent the powers of 10.}
    \begin{threeparttable}
        \begin{tabular}{l d{13.17} d{3.8}}
            \toprule
            \multicolumn{1}{c}{Contribution}              & \multicolumn{1}{c}{$2 \, ^1 P$} & \multicolumn{1}{c}{$2 \, ^3 P$} \\
            \midrule
            $E_{\mathrm{st}}$                             & 4.746\,210                      & -1.582\,070                     \\
            $E_{\mathrm{R1}}$                             & 1.993\,413                      & -20.643\,564                    \\
            $E_{\mathrm{R2}}$                             & 0.002\,086                      & -0.184\,096                     \\
            $m \alpha^6 \ln (\alpha)$~\cite{Drake_1993}   & 0.212\,0                        & 1.64[-7]                        \\
            $E^{(6)}$~\cite{Pachucki_2017, Pachucki_2021} & 8.818                           & -21.833                         \\
            $E^{(7)}$~\cite{Pachucki_2017, Pachucki_2021} & 0.81                            & 2.280                           \\
            $E_{\mathrm{rmdr}}$                           & 2.674                           & 2.857                           \\
            \bottomrule
        \end{tabular}
    \end{threeparttable}
\end{table}

\subsection{Ionization energies}

\begin{figure}
    \centering
    \begin{minipage}[b]{0.37\textwidth} 
        \includegraphics[scale=0.315]{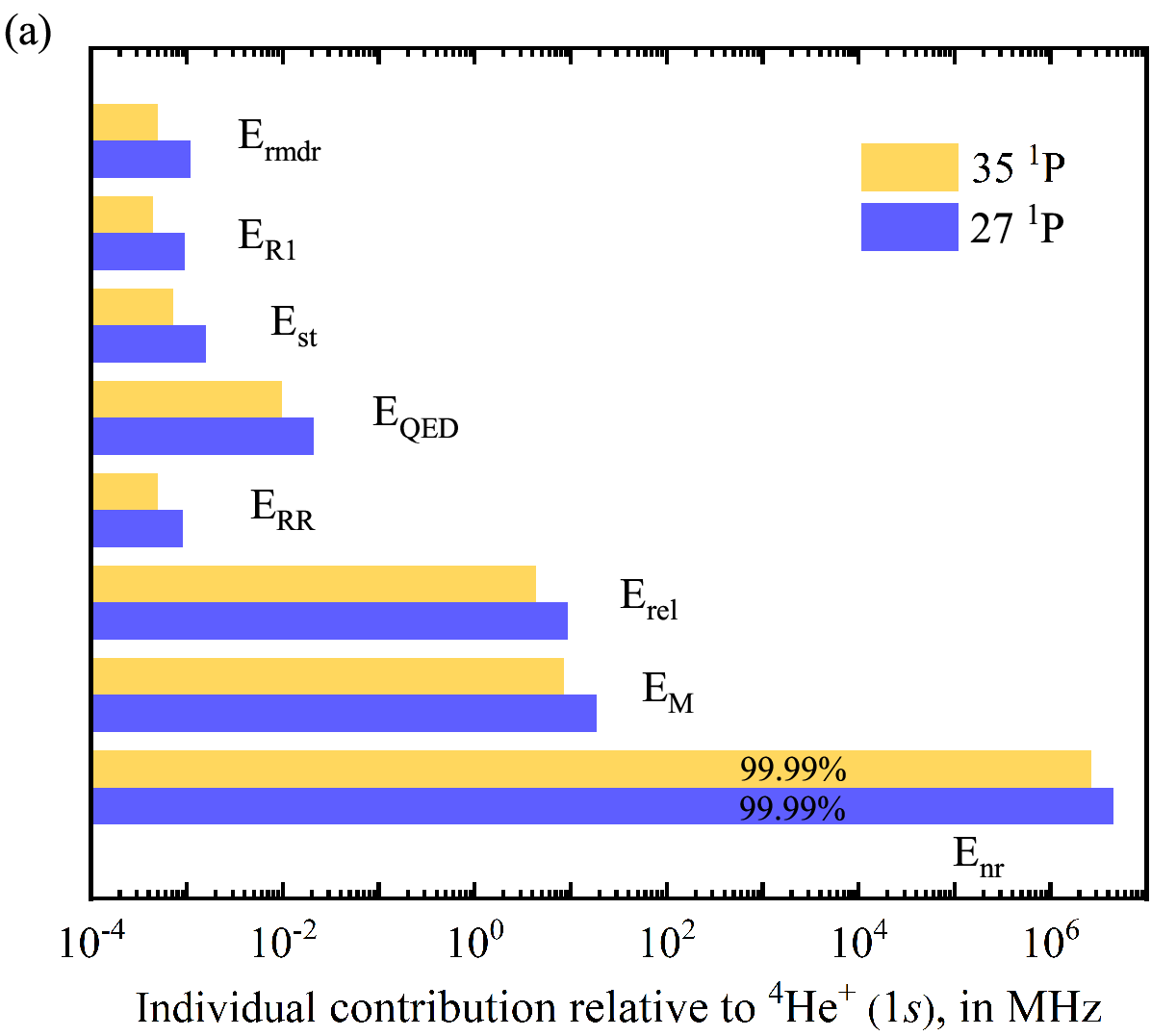}
        \label{fig:Contribution_1P}
    \end{minipage}
    \hfill
    \begin{minipage}[b]{0.37\textwidth}
        \includegraphics[scale=0.315]{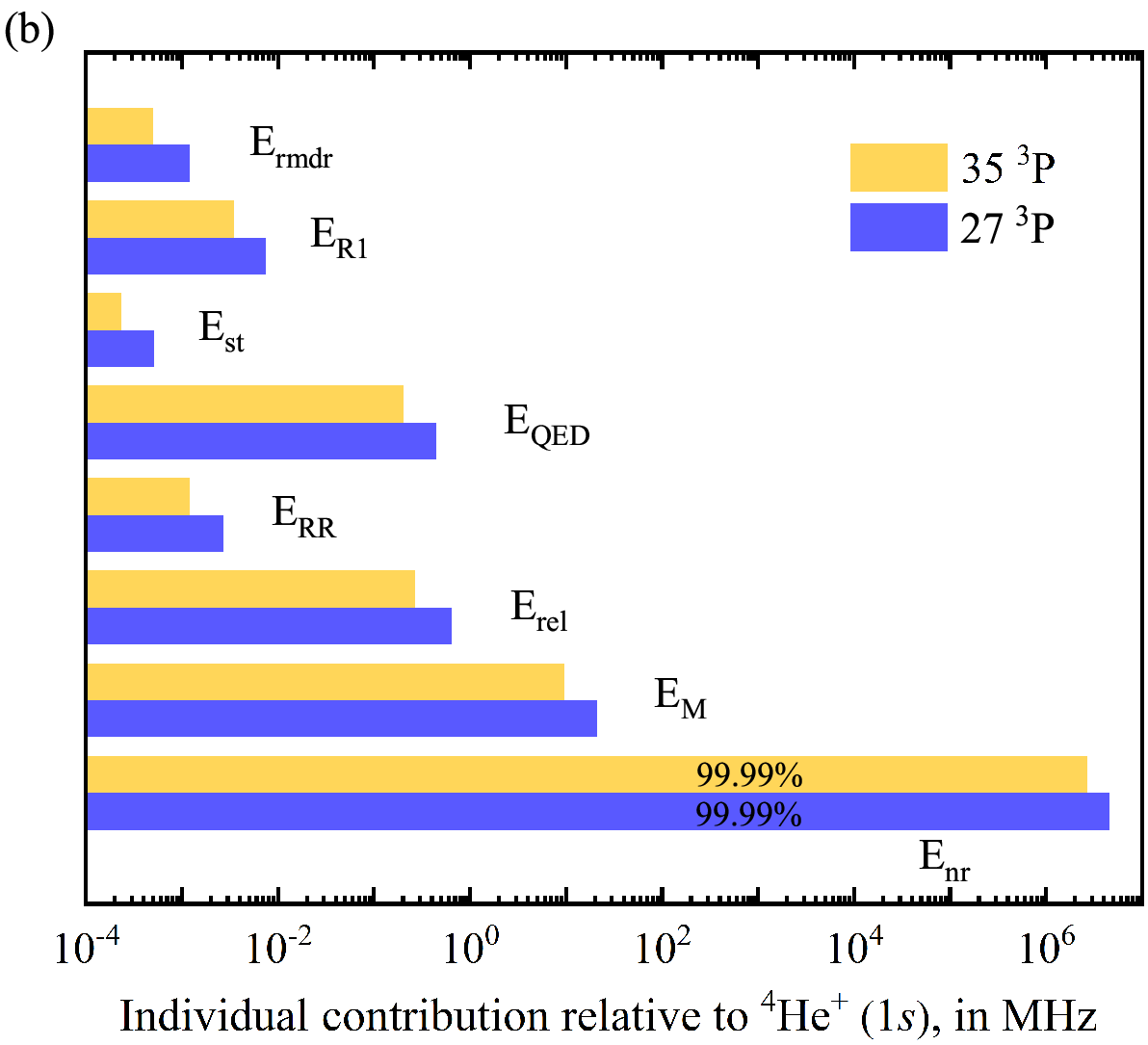}
        \label{fig:Contribution_3P}
    \end{minipage}
    \caption{Percentage contributions to the ionization energies of $^4\mathrm{He}$ relative to $^4 \mathrm{He}^+(1s)$: (a) $35\,^1P$ and $27\,^1P$; (b) $35\,^3P$ and $27\,^3P$. Here $E_{\mathrm{nr}}$, $E_{\mathrm{M}}$, $E_{\mathrm{rel}}$, $E_{\mathrm{RR}}$, $E_{\mathrm{QED}}$, $E_{\mathrm{st}}$, $E_{\mathrm{R1}}$, and $E_{\mathrm{rmdr}}$ denote the nonrelativistic, finite-nuclear-mass, leading relativistic, relativistic recoil, leading QED, singlet-triplet mixing, one-loop radiative, and remainder contributions, respectively. Only the magnitudes of the individual contributions are shown.}
    \label{fig:Contribution_1P_3P}
\end{figure}

To examine the individual contributions to the ionization energies of the $27 \, ^{1,3}P$ and $35 \, ^{1,3}P$ states of $^4 \mathrm{He}$ relative to $^4 \mathrm{He}^+(1s)$, the contributions at different orders are listed in Table~\ref{tab:order_contribution}, and their percentage contributions are displayed in Fig.~\ref{fig:Contribution_1P_3P}. The $E_{R2}$ contribution is omitted from Fig.~\ref{fig:Contribution_1P_3P}. The $^4 \mathrm{He}^+(1s)$ contributions at each order are evaluated according to Ref.~\cite{Salman_2026}. In Table~\ref{tab:order_contribution}, the symbol $(\mu/M)^+$ denotes the finite-nuclear-mass correction including all relevant orders, and the corresponding results obtained with the Hylleraas basis are also presented for comparison.

\begin{table*}[ht]
    \footnotesize
    \caption{\label{tab:order_contribution} Breakdown of the ionization energies of the $27 \, ^1 P$, $27 \, ^3 P$, $35 \, ^1 P$, and $35 \, ^3 P$ states of $^4 \mathrm{He}$ relative to $^4 \mathrm{He}^+ \, (1s)$ (in MHz). Here $E_{\mathrm{nr}}$ is the nonrelativistic contribution, $E_{\mathrm{M}}$ the finite-nuclear-mass correction, $E_{\mathrm{rel}}$ the leading relativistic correction, $E_{\mathrm{RR}}$ the relativistic recoil correction, and $E_{\mathrm{QED}}$ the leading QED correction. The term $E_{\mathrm{st}}$ denotes the singlet-triplet mixing correction, $E_{\mathrm{R1}}$ and $E_{\mathrm{R2}}$ the one-loop and two-loop radiative corrections, respectively, and $E_{\mathrm{rmdr}}$ the remainder contribution from Table~\ref{tab:malpha6_remainder} scaled by a factor of $(2/n)^3$. The notation $(\mu/M)^+$ indicates inclusion of finite nuclear mass effects to all orders. For each contribution, the first line presents the present results and the second line lists Hylleraas reference data~\cite{Drake_2026}. The numbers in parentheses and square brackets are the uncertainties and powers of 10, respectively.}
    \begin{threeparttable}
        \begin{tabular*}{\textwidth}{@{\extracolsep{\fill}} l l d{10.12} d{10.12} d{10.12}  d{10.12}}
            \toprule
            \multicolumn{1}{l}{Contribution} & \multicolumn{1}{l}{Order} & \multicolumn{1}{c}{$27 \, ^1 P$} & \multicolumn{1}{c}{$27 \, ^3 P$} & \multicolumn{1}{c}{$35 \, ^1 P$} & \multicolumn{1}{c}{$35 \, ^3 P$}          \\
            \midrule
            $E_{\mathrm{nr}}$   & $m \alpha^2$               & 4\,508\,153.326\,639(5)   & 4\,535\,100.640\,700(5)   & 2\,683\,360.240(1)        & 2\,695\,724.579(1)          \\
            $\quad$             &                            & 4\,508\,153.326\,638 6    & 4\,535\,100.640\,700\,2   & 2\,683\,360.239\,567      & 2\,695\,724.579\,595        \\
            $E_M$               & $m \alpha^2 \, (\mu/M)^+$  & -18.723\,07(4)            & 21.000\,19(4)             & -8.588(2)                 & 9.634(2)                    \\
            $\quad$             &                            & -18.723\,067\,1           & 21.000\,190\,9            & -8.587\,80(2)             & 9.634\,21(3)                \\
            $E_{\mathrm{rel}}$  & $m \alpha^4$               & 9.369\,8(3)               & -0.643\,2(3)              & 4.329\,1(3)               & -0.268\,0(3)                \\
            $E_{\mathrm{RR}}$   & $m \alpha^4 \, (\mu/M)$    & 0.000\,9(3)               & -0.002\,7(3)              & 0.000\,5(3)               & -0.001\,2(3)                \\
            $E_{\mathrm{QED}}$  & $m \alpha^5$               & -0.021\,25(6)             & 0.447\,15(6)              & -0.009\,77(6)             & 0.204\,89(6)                \\
            $\quad$             &                            & -0.021\,425\,5            & 0.447\,211\,4             & -0.009\,835\,7            & 0.204\,943\,8               \\
            $E_{\mathrm{st}}$   & $m \alpha^6$               & -0.001\,597               & 0.000\,513                & -0.000\,730               & 0.000\,236                  \\
            $E_{\mathrm{R1}}$   & $m \alpha^6$               & -0.000\,958               & 0.007\,550                & -0.000\,440               & 0.003\,460                  \\
            $E_{\mathrm{R2}}$   & $m \alpha^6$               & -3.280\,64(2)[-7]         & 6.733\,265[-5]            & -1.489\,31(2)[-7]         & 3.085\,505[-5]              \\
            $E_{\mathrm{rmdr}}$ &                            & -0.001\,1(11)             & -0.001\,2(12)             & -0.000\,5(5)              & -0.000\,5(5)                \\
            Total               &                            & 4\,508\,143.949\,4(12)    & 4\,535\,121.449\,1(12)    & 2\,683\,355.970\,2(24)    & 2\,695\,734.151\,9(24)      \\
            \bottomrule
        \end{tabular*}
    \end{threeparttable}
\end{table*}

From Fig.~\ref{fig:Contribution_1P_3P}, for both the singlet and triplet series, the percentage contributions of the correction terms generally decrease with increasing $n$, while $E_{\mathrm{nr}}$ remains the dominant contribution throughout. For the $n \, ^1 P$ states, $E_{\mathrm{M}}$ and $E_{\mathrm{rel}}$ have comparable percentage contributions, whereas for the $n \, ^3 P$ states, the contribution of $E_{\mathrm{rel}}$ is clearly smaller than $E_{\mathrm{M}}$. The $E_{\mathrm{QED}}$ contribution is also larger in the triplet states than in the corresponding singlet states. Likewise, $E_{\mathrm{st}}$ and $E_{\mathrm{R1}}$ contribute at comparable levels in the singlet states, whereas for the triplet states, $E_{\mathrm{st}}$ contributes less than $E_{\mathrm{R1}}$. In addition, the $E_{\mathrm{R1}}$ contribution for the singlet states is also smaller than that for the corresponding triplet states.

The present results obtained at each order are in agreement with the Hylleraas results. A small discrepancy is observed in the QED correction, arising from the recoil correction that is included in the Hylleraas results but omitted in the present C-BSBF treatment. For the total ionization energies, the uncertainty of the $27 \, ^{1,3}P$ states is dominated by the estimate of the remainder contribution $E_{\mathrm{rmdr}}$, whereas that for the $35 \, ^{1,3}P$ states is primarily limited by the accuracy of the wave functions. Overall, the uncertainties of the ionization energies are within \num{2.5}~kHz.

The calculated ionization energies of the Rydberg $n \, ^{1,3}P$ states for $n=4$--35 are summarized in Table~\ref{tab:ionization_1P_3P_ST_6_6} in Appendix A. The C-BSBF results achieve kHz-level accuracy, and the corresponding Hylleraas results~\cite{Drake_2026} are also included for comparison. The two approaches are consistent within uncertainties, and together provide reliable benchmark values for the $n \, ^{1,3}P$ states. Furthermore, in order to cover the experimentally measured transitions involving higher principal quantum numbers beyond the range directly accessible by the present \emph{ab initio} calculations, the calculated results are extrapolated using QDT. The corresponding fitted parameters are listed in Appendix B. Based on $E_I(nP)$ values obtained from both the \emph{ab initio} calculations and the quantum-defect extrapolation, the ionization energies of the metastable $2 \, ^1 S$ and $2 \, ^3 S$ states can then be determined in combination with the high-precision experimental transition frequencies~\cite{Clausen_2021, Clausen_4He_2025}.

Using the Rydberg-state ionization energies $E_I(nP)$, the ionization energies of the metastable $2 \, ^1 S$ and $2 \, ^3 S$ states can be determined by combining these values with the high-precision transition frequencies measured by Clausen \emph{et al.}~\cite{Clausen_2021,Clausen_4He_2025}. The C-BSBF \emph{ab initio} calculations are performed for the $n=24$--35 Rydberg states, supplemented by quantum-defect extrapolated values for the $n \, ^1P$ states with $n=40$--102 and the $n \, ^3P$ states with $n=40$--55. The results are listed in Table~\ref{tab:ionization_comparison}. For each $n$, the uncertainty of an individual determination of the $2 \, ^1 S$ and $2 \, ^3 S$ ionization energies is taken as the linear sum of the uncertainty in $E_I(nP)$ and in the transition frequency $\nu_{\mathrm{exp}}(2S - nP)$. Multiple determinations are obtained for each of the $2 \, ^1 S$ and $2 \, ^3 S$ states, which are combined in a statistical analysis to extract the final results. The central values are determined by weighted averaging, while the uncertainties are taken as the root-mean-square (RMS) of the individual uncertainties, scaled by $1/\sqrt{N-1}$, where $N$ denotes the number of measurements.

\begin{figure}
    \centering
    \begin{minipage}[b]{0.40\textwidth} 
        \includegraphics[scale=0.315]{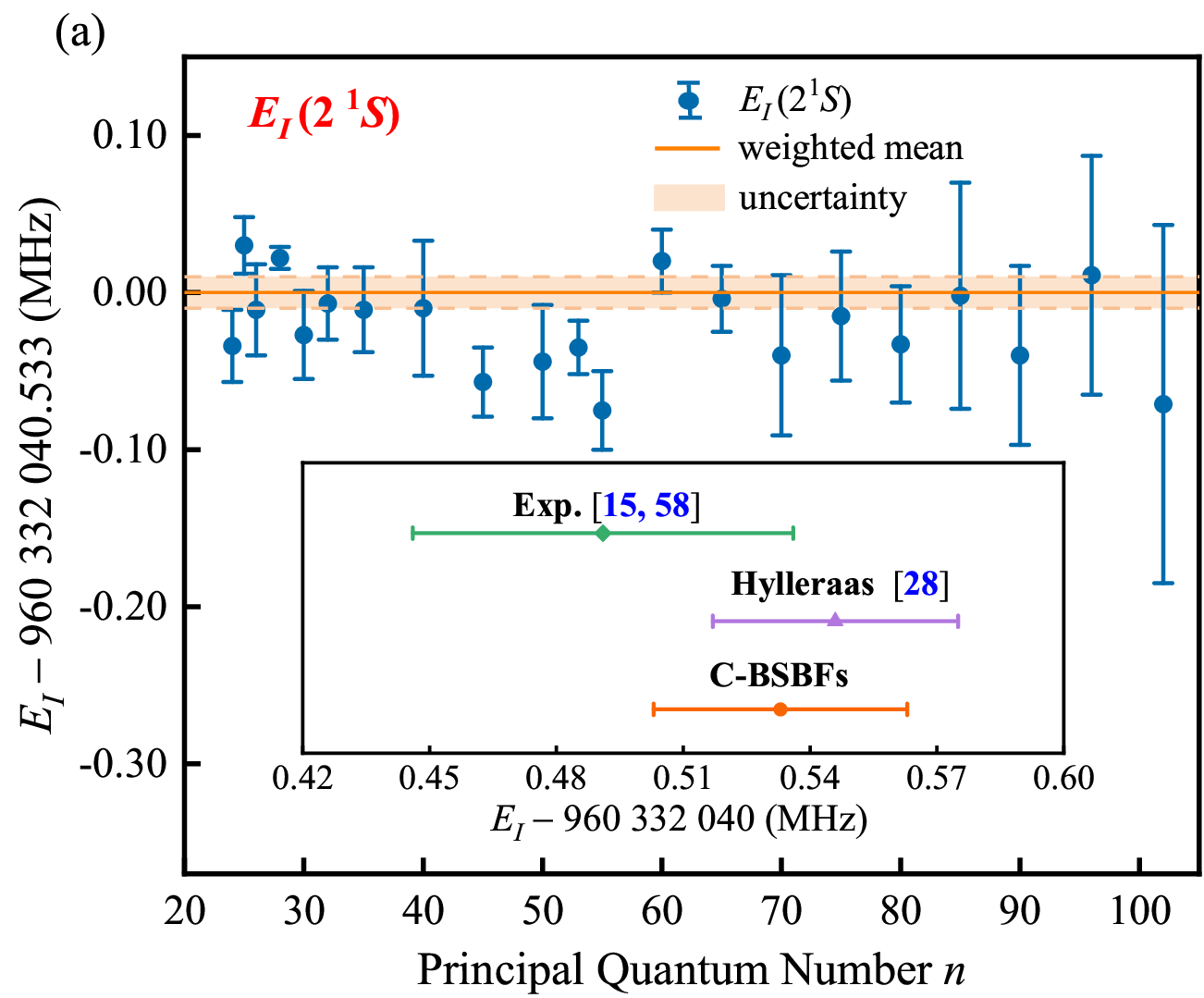}
        \label{fig:Ionization_1S}
    \end{minipage}
    \hfill
    \begin{minipage}[b]{0.40\textwidth}
        \includegraphics[scale=0.315]{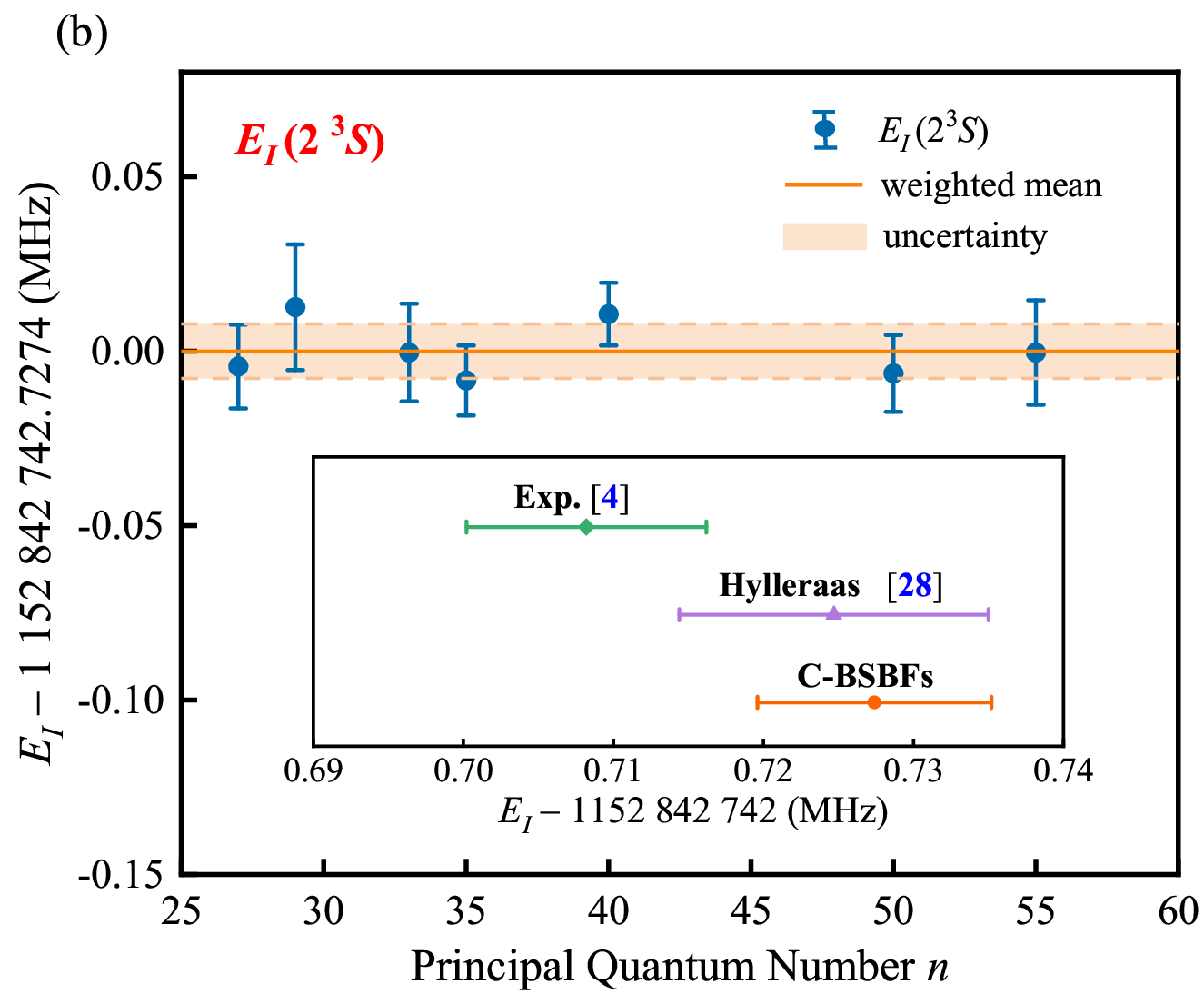}
        \label{fig:Ionization_3S}
    \end{minipage}
    \caption{Ionization energies of (a) the $2 \, ^{1} S$ and (b) the $2 \, ^{3} S$ states for $^4 \mathrm{He}$, in MHz. The values marked in blue are the ionization energies determined at different principal quantum numbers $n$, as listed in the last column of Table~\ref{tab:ionization_comparison}, with error bars representing the corresponding uncertainties. The solid orange lines denote weighted-average central values of the ionization energies, and the shaded bands represent the statistical uncertainties of the averages, 10~kHz for (a) and 5.3~kHz for (b), respectively. The inset shows comparisons between theoretical and experimental results.}
    \label{fig:Ionization_1S_3S}
\end{figure}

Table~\ref{tab:ionization_comparison} lists the metastable-state ionization energies extracted from different Rydberg states, together with the corresponding weighted averages and uncertainties. Based on the tabulated data, Figs.~\ref{fig:Ionization_1S_3S}(a) and \ref{fig:Ionization_1S_3S}(b) further display the individual Rydberg-state results and their weighted averages. The orange solid lines denote the central values of the weighted averages and the orange shaded bands denote the corresponding uncertainties. The resulting C-BSBF ionization energy for the $2 \, ^1S$ state is \num{960332040.533(10)}$_\mathrm{stat}(20)_\mathrm{sys}$~MHz. The corresponding value for the $2 \, ^3S$ state, derived from the ionization energies extracted via the $n \, ^3P$ Rydberg states, is \num{1152842742.7274(53)}$_\mathrm{stat}(25)_\mathrm{sys}$~MHz. The first uncertainty denotes the statistical contribution from the individual Rydberg-state determinations, while the second represents the systematic uncertainty from the experimental transition frequencies. For the weighted-average ionization energies, the statistical uncertainties are mainly determined by the uncertainties in $\nu_{\mathrm{exp}}(2S \rightarrow nP)$, giving \num{10}~kHz for the $2 \, ^1S$ state and \num{5.3}~kHz for the $2 \, ^3S$ state. The smaller uncertainty in the $2 \, ^3S$ result reflects the higher precision of the corresponding transition-frequency measurements.

In addition, an independent determination of the $2 \, ^3 S$ ionization energy can be derived from $E_I(2 \, ^1 S)$ using the high-precision transition frequency $\nu(2 \, ^1S \rightarrow 2 \, ^3S)=\num{192510702.14872(20)}$~MHz~\cite{Rengelink_2018}. This yields $E_I(2 \, ^3S)=\num{1152842742.682(10)}_\mathrm{stat}(20)_\mathrm{sys}$~MHz, which is lower by about \num{45.4}~kHz than the value obtained from the $n \, ^3P$ Rydberg-state determinations. This indirect determination provides an independent reference for the $2 \, ^3S$ ionization energy. In this work, the directly determined value from the $n \, ^3P$ Rydberg-state determinations, $E_I(2 \, ^3S)=\num{1152842742.7274(53)}_\mathrm{stat}(25)_\mathrm{sys}$~MHz, is adopted as the C-BSBF ionization energy of the $2 \, ^3S$ state.

As shown in Table~\ref{tab:ionization_comparison}, the present C-BSBF results can be directly compared with the Hylleraas calculations~\cite{Drake_2026}. It should be noted that the Hylleraas results were derived from explicitly calculated Rydberg states over the $n=24$--35 interval for the singlet series and the $n=27$--35 interval for the triplet series~\cite{Drake_2026}. Carrying out the present C-BSBF calculation over the same Rydberg-state intervals gives ionization energies of \num{960332040.546(9)}$_\mathrm{stat}(20)_\mathrm{sys}$~MHz for the $2 \, ^1S$ state and \num{1152842742.7243(80)}$_\mathrm{stat}(25)_\mathrm{sys}$ MHz for the $2 \, ^3S$ state. Over these Rydberg-state intervals, the C-BSBF and Hylleraas determinations show good agreement. This agreement provides a reference for assessing the effect of the additional higher-$n$ Rydberg-state data included in the C-BSBF weighted averages. Including higher-$n$ Rydberg-state data shifts the C-BSBF weighted averages by about \num{13}~kHz for the $2 \, ^1S$ state and about \num{3.1}~kHz for the $2 \, ^3S$ state. These shifts suggest that higher-$n$ Rydberg-state data provide complementary constraints on the extraction of metastable-state ionization energies, and that their inclusion helps assess the stability of the weighted averages.

The comparison with experiment is also presented in Table~\ref{tab:ionization_comparison}, with the relative positions of the theoretical and experimental values further illustrated in the insets of Figs.~\ref{fig:Ionization_1S_3S}(a) and \ref{fig:Ionization_1S_3S}(b). The ionization energies obtained from the two theoretical approaches are in good agreement. For the C-BSBF results, the $2 \, ^1S$ ionization energy agrees with the experimental value obtained from the Rydberg-series extrapolation~\cite{Clausen_2021, Clausen_2025}. For the $2 \, ^3S$ state, comparison with the corresponding experimental value yields a small difference of \num{0.019(10)}~MHz. In evaluating these deviations, the statistical and systematic uncertainties are added linearly. The current uncertainty in the extracted metastable-state ionization energies is mainly limited by the experimental transition frequencies $\nu_{\mathrm{exp}}(2S \rightarrow nP)$. Further improvement would require both extending the data set to additional $2S \rightarrow nP$ Rydberg transitions and reducing the uncertainties of the individual transition-frequency measurements, for example through better control of Doppler-related effects and field-induced shifts as well as improved frequency calibration. These improvements would allow the metastable-state ionization energies to be determined with higher precision and would enable more stringent tests of theoretical calculations.

\begin{table*}[ht]
    \footnotesize
    \caption{\label{tab:ionization_comparison} Ionization energies of the metastable $2 \, ^{1} S$ and $2 \, ^{3} S$ states of $^4 \mathrm{He}$ (in MHz), obtained by combining Rydberg-state ionization energies with the corresponding experimental transition frequencies. The ionization energies $E_I(n P)$ for $n=24$--35 are obtained from the present C-BSBF \emph{ab initio} calculations, whereas those for $n=40$--102 are derived from the quantum-defect fit based on Eqs.~\eqref{eq:ionization_energy} and \eqref{eq:quantum_defect}. $\nu_{\mathrm{exp}}$($2S \rightarrow nP$) denotes the experimental transition frequencies, and $E_I(2 \, ^{1}S)$ and $E_I(2 \, ^{3}S)$ are the ionization energies of the corresponding metastable states. ``Difference'' represents the deviation between the C-BSBF weighted-average result and the experimental data~\cite{Clausen_2021,Clausen_4He_2025}. The Hylleraas values~\cite{Drake_2026} are also listed for comparison. }
    \begin{threeparttable}
        \begin{tabular*}{\textwidth}{@{\extracolsep{\fill}} l d{8.9} d{10.8} d{10.16} d{10.17}}
            \toprule
            \multicolumn{1}{l}{$n$} & \multicolumn{1}{c}{$E_I(n P)$}  & \multicolumn{1}{c}{$\nu_{\mathrm{exp}}$($2S \rightarrow nP$)}  & \multicolumn{1}{c}{$E_I(2 \, ^1 S)$} & \multicolumn{1}{c}{$E_I(2 \, ^3 S)$} \\
            \midrule
            \multicolumn{5}{c}{$^1 P$~\tnote{a}}                                                                                                                             \\
            24            & 5\,704\,980.348\,2(16)   & 954\,627\,060.151(22)   & 960\,332\,040.499(23)            & 1\,152\,842\,742.648(23)         \\
            25            & 5\,257\,921.948\,6(14)   & 955\,074\,118.614(17)   & 960\,332\,040.563(18)            & 1\,152\,842\,742.712(18)         \\
            26            & 4\,861\,425.438\,0(13)   & 955\,470\,615.084(28)   & 960\,332\,040.522(29)            & 1\,152\,842\,742.671(29)         \\
            28            & 4\,192\,018.126\,3(10)   & 956\,140\,022.429(6)    & 960\,332\,040.555(7)             & 1\,152\,842\,742.704(7)          \\
            30            & 3\,651\,924.176\,0(9)    & 956\,680\,116.330(27)   & 960\,332\,040.506(28)            & 1\,152\,842\,742.655(28)         \\
            32            & 3\,209\,861.007\,0(8)    & 957\,122\,179.519(22)   & 960\,332\,040.526(23)            & 1\,152\,842\,742.675(23)         \\
            35            & 2\,683\,355.970\,2(24)   & 957\,648\,684.552(25)   & 960\,332\,040.522(27)            & 1\,152\,842\,742.671(27)         \\
            40            & 2\,054\,622.318\,8(3)    & 958\,277\,418.204(42)   & 960\,332\,040.523(43)            & 1\,152\,842\,742.672(43)         \\
            45            & 1\,623\,514.662\,1(2)    & 958\,708\,525.814(21)   & 960\,332\,040.476(22)            & 1\,152\,842\,742.625(22)         \\
            50            & 1\,315\,117.770\,8(2)    & 959\,016\,922.718(35)   & 960\,332\,040.489(36)            & 1\,152\,842\,742.638(36)         \\
            53            & 1\,170\,482.283\,2(1)    & 959\,161\,558.215(16)   & 960\,332\,040.498(17)            & 1\,152\,842\,742.647(17)         \\
            55            & 1\,086\,922.138\,9(1)    & 959\,245\,118.319(24)   & 960\,332\,040.458(25)            & 1\,152\,842\,742.607(25)         \\
            60            & 913\,350.099\,9(1)       & 959\,418\,690.453(19)   & 960\,332\,040.553(20)            & 1\,152\,842\,742.702(20)         \\
            65            & 778\,263.352\,56(8)      & 959\,553\,777.176(21)   & 960\,332\,040.529(21)            & 1\,152\,842\,742.678(21)         \\
            70            & 671\,071.502\,83(6)      & 959\,660\,968.990(51)   & 960\,332\,040.493(51)            & 1\,152\,842\,742.642(51)         \\
            75            & 584\,591.355\,86(5)      & 959\,747\,449.162(41)   & 960\,332\,040.518(41)            & 1\,152\,842\,742.667(41)         \\
            80            & 513\,811.389\,51(4)      & 959\,818\,229.110(37)   & 960\,332\,040.500(37)            & 1\,152\,842\,742.649(37)         \\
            85            & 455\,149.008\,62(4)      & 959\,876\,891.522(72)   & 960\,332\,040.531(72)            & 1\,152\,842\,742.680(72)         \\
            90            & 405\,988.118\,87(3)      & 959\,926\,052.374(57)   & 960\,332\,040.493(57)            & 1\,152\,842\,742.642(57)         \\
            96            & 356\,831.510\,65(2)      & 959\,975\,209.033(76)   & 960\,332\,040.544(76)            & 1\,152\,842\,742.693(76)         \\
            102           & 316\,090.746\,01(2)      & 960\,015\,949.716(114)  & 960\,332\,040.462(114)           & 1\,152\,842\,742.611(114)        \\
            Weighted avg. &                          &                         & 960\,332\,040.533(10)_\mathrm{stat}(20)_\mathrm{sys}  & 1\,152\,842\,742.682(10)_\mathrm{stat}(20)_\mathrm{sys}  \\
            Hylleraas~\cite{Drake_2026} &            &                         & 960\,332\,040.546(9)_\mathrm{stat}(20)_\mathrm{sys}   & 1\,152\,842\,742.694(10)_\mathrm{stat}(20)_\mathrm{sys}  \\
            Exp.~\cite{Clausen_2021, Clausen_2025}   &            &            & 960\,332\,040.501(25)_\mathrm{stat}(20)_\mathrm{sys}  & 1\,152\,842\,742.650(25)_\mathrm{stat}(20)_\mathrm{sys}  \\
            Difference    &                          &                         & 0.032(47)                                             & 0.032(47)   \\
            \multicolumn{5}{c}{ }                                                                                                                    \\
            \multicolumn{5}{c}{$^3 P$}                                                                                                               \\
            27            & 4\,535\,121.449\,1(12)   & 1\,148\,307\,621.274(11)   &                               & 1\,152\,842\,742.723(12)         \\
            29            & 3\,929\,783.191\,4(10)   & 1\,148\,912\,959.549(17)   &                               & 1\,152\,842\,742.740(18)         \\
            33            & 3\,033\,110.360\,0(8)    & 1\,149\,809\,632.367(13)   &                               & 1\,152\,842\,742.727(14)         \\
            35            & 2\,695\,734.151\,9(24)   & 1\,150\,147\,008.567(8)    &                               & 1\,152\,842\,742.719(10)         \\
            40            & 2\,062\,912.749\,5(4)    & 1\,150\,779\,829.988(8)    &                               & 1\,152\,842\,742.738(9)          \\
            50            & 1\,319\,360.989\,9(2)    & 1\,151\,523\,381.731(10)   &                               & 1\,152\,842\,742.721(11)         \\
            55            & 1\,090\,109.714\,7(1)    & 1\,151\,752\,633.012(14)   &                               & 1\,152\,842\,742.727(15)         \\
            Weighted avg. &                          &                            &                               & 1\,152\,842\,742.727\,4(53)_\mathrm{stat}(25)_\mathrm{sys} \\
            Hylleraas~\cite{Drake_2026}  &           &                            &                               & 1\,152\,842\,742.724\,7(78)_\mathrm{stat}(25)_\mathrm{sys} \\
            Exp.~\cite{Clausen_4He_2025} &           &                            &                               & 1\,152\,842\,742.708\,2(55)_\mathrm{stat}(25)_\mathrm{sys} \\
            Difference    &                          &                            &                               & 0.019(10)                        \\
            \bottomrule
        \end{tabular*}
        \begin{tablenotes}
            \item[a] Experimental $\nu_{\mathrm{exp}}(2S$--$nP)$ values for the $^1P$ series are taken from Ref.~\cite{Clausen_2025} and private communication with F. Merkt.
        \end{tablenotes}
    \end{threeparttable}
\end{table*}

\section{Conclusion}

In this work, the correlated B-spline basis function (C-BSBF) method was extended for high-accuracy calculations of the ionization energies in the Rydberg $^{1,3} P$ states of $^4 \mathrm{He}$. Using a unified basis set, corrections of orders $m \alpha^4$, including relativistic recoil contributions, and $m \alpha^5$, as well as singlet-triplet mixing contributions and one- and two-loop radiative terms of order $m \alpha^6$ were evaluated explicitly. The remaining nonradiative contributions of orders $m \alpha^6$ and $m \alpha^7$ were estimated using a $1/n^3$ scaling extrapolation from $2 \, ^1 P$ and $2 \, ^3 P$ states~\cite{Pachucki_2006_energy, Pachucki_2021}.

For principal quantum numbers $n = 24$--35, the \emph{ab initio} calculations yield ionization energies of the $n \, ^{1,3} P$ states with uncertainties at the kHz level. The C-BSBF results are consistent with the corresponding Hylleraas values~\cite{Drake_2026}, confirming the reliability of the Rydberg-state calculations. Based on these results, quantum-defect parameters were determined, enabling reliable extrapolation to higher-$n$ states. The fitted $n \, ^{1,3} P$ ionization energies for $n=40$--102 are obtained with a precision better than 100 Hz. By combining the ionization energies with the high-precision experimental transition frequencies $\nu_{\mathrm{exp}}(2S \rightarrow nP)$, the ionization energies of the metastable $2 \, ^1 S$ and $2 \, ^3 S$ states were determined to be \num{960332040.533(10)}$_\mathrm{stat}(20)_\mathrm{sys}$~MHz and \num{1152842742.7274(53)}$_\mathrm{stat}(25)_\mathrm{sys}$~MHz, respectively. The present C-BSBF results are consistent with the high-accuracy Hylleraas calculations~\cite{Drake_2026}, providing an independent cross-validation of the extracted metastable-state ionization energies. Compared with the experimental ionization energies obtained from Rydberg-series extrapolations~\cite{Clausen_2021, Clausen_4He_2025, Clausen_2025}, the C-BSBF value for the $2 \, ^1 S$ state is consistent with experiment. For the $2 \, ^3 S$ state, a small difference of \num{0.019(10)}~MHz remains relative to the extrapolated experimental value. Further improvement in the $2S \rightarrow nP$ experimental transition frequencies, together with high-precision measurements of additional Rydberg transitions, will therefore be important for more stringent tests of theoretical calculations.

Owing to the flexibility of the B-spline basis construction, the C-BSBF method can be extended, within the same basis-set configuration and parameters, to high-accuracy calculations of the Rydberg $^{1,3} S$ and $^{1,3} D$ states. This capability also enables investigations of transition frequencies involving low-lying states, several of which exhibit reported discrepancies between experiment and theory. The most notable cases are the $2 \, ^3S$--$3 \, ^3D_1$ and $2 \, ^3P_0$--$3 \, ^3D_1$ transition frequencies, whose experimental values differ from the theoretical predictions by about $6\sigma$ and $15\sigma$, respectively~\cite{Pachucki_2021}. Recently, Doppler-free two-photon spectroscopy of the $2 \, ^3S_1$--$8 \, ^3D_{1,2,3}$ transitions using a power-enhanced cavity~\cite{Wumuhao_2024}, followed by the absolute frequency measurement of the $2 \, ^3S_1$--$8 \, ^3D_{1}$ line~\cite{Wumuhao_2025}, has demonstrated both the feasibility of precision measurements involving higher $n \, ^3D$ states and their potential for determining the $2 \, ^3S$ ionization energy. These results suggest that similar measurements can be extended to higher-$n$ states in the $2 \, ^3S_1$--$n \, ^3D$ Rydberg series. Calculating the ionization energies of the corresponding $n \, ^3D$ Rydberg states using the C-BSBF method, together with the $S$-state results, would further test the applicability of the method beyond the $P$ series. Combining these theoretical results with future experimental transition frequencies would provide additional information for clarifying the discrepancies between theory and experiment in precision helium spectroscopy.

\begin{acknowledgments}
    The authors thank X.-Q. Qi for his contribution in the early stages of this work, F. Merkt for providing the original experimental data, G. W. F. Drake and Z.-C. Yan for helpful discussions on ionization energy calculations. This work is supported by the National Natural Science Foundation of China under Grants No. 12393821, No. 12274423, and No. 12274417, by the Chinese Academy of Sciences Project for Young Scientists in Basic Research under Grant No. YSBR-055, and by the Pioneer Research Project for Basic and Interdisciplinary Frontiers of Chinese Academy of Sciences under Grants No. XDB0920101 and XDB0920100. Theoretical calculations were done on the APM-Theoretical Computing Cluster (APM-TCC). 
\end{acknowledgments}

\section*{DATA AVAILABILITY}
The data supporting the findings of this article have been tabulated within the article. Additional metadata are available from the corresponding author upon request.

\section*{Appendix A: Ionization Energies of $^{1,3}P$ States in $^4\mathrm{He}$}

This appendix presents the calculated ionization energies of the $n \, ^{1,3} P$ states in $^4 \mathrm{He}$ for principal quantum numbers $n=4$--35. Table~\ref{tab:ionization_1P_3P_ST_6_6} lists both the present C-BSBF ionization frequencies and the recent high-accuracy calculations of Drake \emph{et al.}~\cite{Drake_2026}, enabling a direct comparison between two sets of results obtained using independent theoretical approaches. The C-BSBF results generally achieve kHz-level accuracy. The uncertainties are relatively larger for the lower-n states and decrease toward the higher Rydberg states. This behavior mainly reflects the larger influence of the uncalculated nonradiative contributions of orders $m \alpha^6$ and $m \alpha^7$ on the lower-n states, whereas these contributions become progressively smaller along the Rydberg series, approximately following the $1/n^3$ scaling. The two data sets are consistent within uncertainties throughout the range of $n$ considered, providing a reliable cross-validation between the two \emph{ab initio} approaches and further supporting the accuracy of the present calculations.

\begin{table*}[ht]
    \footnotesize
    \caption{\label{tab:ionization_1P_3P_ST_6_6} The ionization energies of the $n \,^{1,3} P$ states ($n=4$--35) of $^4 \mathrm{He}$ relative to $^4 \mathrm{He}^+ \, (1s)$ (in MHz). The corresponding Hylleraas results~\cite{Drake_2026} are also listed for comparison.}
    \begin{threeparttable}
        \begin{tabular*}{\textwidth}{@{\extracolsep{\fill}} l d{13.10} d{10.10} d{13.10} d{10.10}}
            \toprule
            \multirow{2}{*}{$n$} & \multicolumn{2}{c}{$E_I(n \, ^1 P)$} & \multicolumn{2}{c}{$E_I(n \, ^3 P)$}                                      \\
            \cmidrule(lr){2-3} \cmidrule(lr){4-5}
            & \multicolumn{1}{c}{C-BSBF}           & \multicolumn{1}{c}{Hylleraas~\cite{Drake_2026}} & \multicolumn{1}{c}{C-BSBF} & \multicolumn{1}{c}{Hylleraas~\cite{Drake_2026}} \\
            \midrule
            4                    & 204\,397\,210.39(33)                 & 204\,397\,210.37(33)            & 212\,661\,130.05(36)           & 212\,661\,130.22(29)           \\
            5                    & 130\,955\,541.66(17)                 & 130\,955\,541.64(17)            & 135\,204\,995.70(18)           & 135\,204\,995.78(15)           \\
            6                    & 91\,009\,810.422(99)                 & 91\,009\,810.41(10)             & 93\,472\,929.82(11)            & 93\,472\,929.86(8)             \\
            7                    & 66\,901\,127.442(62)                 & 66\,901\,127.43(6)              & 68\,453\,141.708(67)           & 68\,453\,141.74(5)             \\
            8                    & 51\,242\,587.328(42)                 & 51\,242\,587.32(4)              & 52\,282\,461.841(45)           & 52\,282\,461.860(35)           \\
            9                    & 40\,501\,246.346(29)                 & 40\,501\,246.341(29)            & 41\,231\,541.898(31)           & 41\,231\,541.911(24)           \\
            10                   & 32\,814\,665.279(21)                 & 32\,814\,665.275(21)            & 33\,346\,972.316(23)           & 33\,346\,972.326(18)           \\
            11                   & 27\,125\,436.773(16)                 & 27\,125\,436.770(16)            & 27\,525\,292.378(17)           & 27\,525\,292.385(13)           \\
            12                   & 22\,797\,031.712(12)                 & 22\,797\,031.709(12)            & 23\,104\,960.768(13)           & 23\,104\,960.774(10)           \\
            13                   & 19\,427\,674.897\,6(97)              & 19\,427\,674.895(10)            & 19\,669\,820.897(10)           & 19\,669\,820.901(8)            \\
            14                   & 16\,753\,624.159\,4(78)              & 16\,753\,624.158(8)             & 16\,947\,462.114\,2(83)        & 16\,947\,462.118(6)            \\
            15                   & 14\,595\,939.725\,8(64)              & 14\,595\,939.724(6)             & 14\,753\,507.975\,7(68)        & 14\,753\,507.979(5)            \\
            16                   & 12\,829\,749.820\,1(52)              & 12\,829\,749.819(5)             & 12\,959\,559.374\,9(56)        & 12\,959\,559.377\,3(37)        \\
            17                   & 11\,365\,768.213\,4(44)              & 11\,365\,768.212(4)             & 11\,473\,973.543\,5(47)        & 11\,473\,973.545\,4(29)        \\
            18                   & 10\,138\,783.841\,9(37)              & 10\,138\,783.841\,1(36)         & 10\,229\,924.320\,2(39)        & 10\,229\,924.321\,9(23)        \\
            19                   & 9\,100\,271.006\,1(31)               & 9\,100\,271.005\,4(31)          & 9\,177\,753.892\,2(34)         & 9\,177\,753.893\,6(19)         \\
            20                   & 8\,213\,516.624\,3(27)               & 8\,213\,516.623\,7(27)          & 8\,279\,939.622\,9(29)         & 8\,279\,939.624\,3(15)         \\
            21                   & 7\,450\,330.349\,8(23)               & 7\,450\,330.349\,3(23)          & 7\,507\,701.857\,3(25)         & 7\,507\,701.858\,8(12)         \\
            22                   & 6\,788\,776.068\,3(20)               & 6\,788\,776.067\,8(20)          & 6\,838\,668.572\,6(22)         & 6\,838\,668.573\,6(12)         \\
            23                   & 6\,211\,577.817\,8(18)               & 6\,211\,577.817\,4(17)          & 6\,255\,236.666\,2(19)         & 6\,255\,236.667\,2(11)         \\
            24                   & 5\,704\,980.348\,2(16)               & 5\,704\,980.347\,7(15)          & 5\,743\,402.127\,6(17)         & 5\,743\,402.128\,4(11)         \\
            25                   & 5\,257\,921.948\,6(14)               & 5\,257\,921.948\,3(14)          & 5\,291\,911.804\,7(15)         & 5\,291\,911.805\,5(11)         \\
            26                   & 4\,861\,425.438\,0(13)               & 4\,861\,425.437\,7(12)          & 4\,891\,639.559\,6(14)         & 4\,891\,639.560\,3(10)         \\
            27                   & 4\,508\,143.949\,4(12)               & 4\,508\,143.949\,1(11)          & 4\,535\,121.449\,1(12)         & 4\,535\,121.449\,8(10)         \\
            28                   & 4\,192\,018.126\,3(10)               & 4\,192\,018.126\,0(10)          & 4\,216\,205.237\,2(11)         & 4\,216\,205.237\,9(10)         \\
            29                   & 3\,908\,014.558\,5(10)               & 3\,908\,014.558\,2(9)           & 3\,929\,783.191\,4(10)         & 3\,929\,783.191\,9(10)         \\
            30                   & 3\,651\,924.176\,0(9)                & 3\,651\,924.175\,9(8)           & 3\,671\,586.292\,0(9)          & 3\,671\,586.292\,6(10)         \\
            31                   & 3\,420\,205.394\,6(8)                & 3\,420\,205.394\,3(7)           & 3\,438\,024.236\,1(9)          & 3\,438\,024.236\,6(9)          \\
            32                   & 3\,209\,861.007\,0(8)                & 3\,209\,861.006\,7(6)           & 3\,226\,059.950\,3(8)          & 3\,226\,059.950\,8(9)          \\
            33                   & 3\,018\,340.774\,4(8)                & 3\,018\,340.774\,1(6)           & 3\,033\,110.360\,0(8)          & 3\,033\,110.360\,3(9)          \\
            34                   & 2\,843\,463.761\,1(11)               & 2\,843\,463.760\,8(5)           & 2\,856\,967.318\,8(12)         & 2\,856\,967.319\,4(9)          \\
            35                   & 2\,683\,355.970\,2(24)               & 2\,683\,355.969\,7(5)           & 2\,695\,734.151\,9(24)         & 2\,695\,734.153\,0(9)          \\
            \bottomrule
        \end{tabular*}
    \end{threeparttable}
\end{table*}

\section*{Appendix B: Quantum Defect Extrapolations}

Based on the present \emph{ab initio} ionization energies, the corresponding quantum-defect parameters are determined by fitting the ionization energies for $n=4$--33 to~\cite{Drake_1994, Drake_2023springer}
\begin{align}
    E_I(n \, ^{1,3} P) = \frac{R_{\mathrm{He}} c}{\left[ n - \delta(n) \right]^2} , \label{eq:ionization_energy}
\end{align}
where the $n$-dependent quantum defect $\delta(n)$ is represented by the Ritz expansion~\cite{Drake_1994, Drake_1999, Clausen_4He_2025}
\begin{align}
    \delta(n)
     & = \delta_0 + \frac{\delta_2}{\left[ n - \delta(n) \right]^2} + \frac{\delta_4}{\left[ n - \delta(n) \right]^4} \nonumber                       \\
     & \quad + \frac{\delta_6}{\left[ n - \delta(n) \right]^6} + \frac{\delta_8}{\left[ n - \delta(n) \right]^8} + \cdots , \label{eq:quantum_defect}
\end{align}
where $E_I(n \, ^{1,3} P)$ denotes the ionization energy, $R_{\mathrm{He}} = \mu R_{\infty}$ is the mass-corrected Rydberg constant, and $c$ is the speed of light in vacuum. The coefficient $\delta_0$, $\delta_{2}$, $\delta_{4}$, $\delta_{6}$, and $\delta_{8}$ describe the energy dependence of the quantum defect.

In the present calculations, the ionization energies contain relativistic and mass-polarization contributions that are not described by the standard Ritz form. If retained in the ionization energies used for the fit, these contributions would be absorbed into the fitted quantum-defect parameters and could appear as apparent odd-order terms in the expansion. Following Ref.~\cite{Drake_2023springer}, they are therefore subtracted from the calculated ionization energies before the quantum-defect fit is performed:
\begin{align}
    E_I^{\prime}(n \, ^{1,3} P)
     & = E_I(n \, ^{1,3} P)
    - R_{\mathrm{He}} c \bigg\lbrace \frac{-3 \alpha^2 \left( Z - 1 \right)^4}{4 n^4}  \nonumber                                                              \\
     & \quad + \left( \frac{\mu}{M_0} \right)^2 \frac{\left( Z - 1 \right)^2}{n^2} \left[ 1 + \frac{5}{6} \left( \alpha Z \right)^2 \right] \bigg\rbrace \, ,
\end{align}
with $Z=2$ for helium. The quantum-defect parameters are then determined from the corrected ionization energies $E_I^{\prime}(n \, ^{1,3} P)$. As discussed in the recent work of Drake and Bondy~\cite{Drake_2026_latest}, the treatment is equivalent through order $(\mu/M_0)^2$ to using the He$^+$-core Rydberg constant, in which the leading second-order mass-polarization contribution is absorbed into the Rydberg constant. The difference between the two descriptions starts at order $(\mu/M_0)^4$.

The obtained coefficients are listed in Table~\ref{tab:quantum_defect_coefficient}. The ionization energies used to determine these coefficients do not include the remainder contribution $E_{\mathrm{rmdr}}$. This contribution is added subsequently when the fitted ionization energies are evaluated for comparison with the corresponding \emph{ab initio} results. The quantum-defect parameters can also be determined from high-precision spectroscopic measurements, which provide an experimental reference for the present theoretical determination~\cite{Clausen_2021, Clausen_4He_2025}. To assess the reliability of the fit, the fitted results are extrapolated to $n=34$ and 35 and compared with the corresponding \emph{ab initio} calculations. For example, the fitted values for the $35 \, ^1 P$ and $35 \, ^3 P$ states are \num{2683355.97038}~MHz and \num{2695734.15321}~MHz, respectively, which are consistent with the \emph{ab initio} results without including the evaluation of $E_{\mathrm{rmdr}}$, namely \num{2683355.9707(23)}~MHz and \num{2695734.1524(23)}~MHz. The truncation uncertainty of the Ritz expansion is also estimated by repeating the fit with the expansion extended through the $\delta_{10}/[n-\delta(n)]^{10}$ term. The resulting changes in the fitted ionization energies are less than 100 Hz and are therefore negligible at the few kHz precision reported here. These comparisons confirm that the fitted quantum-defect expansion provides a reliable description of the Rydberg series.

\begin{table}[ht]
    \footnotesize
    \caption{\label{tab:quantum_defect_coefficient} Quantum defects for the total ionization energies of $^4 \mathrm{He}$. The numbers in parentheses represent the fitting uncertainties.}
    \begin{threeparttable}
        \begin{tabular}{l d{10.17} d{10.15}}
            \toprule
            \multicolumn{1}{c}{$\delta_i$} & \multicolumn{1}{c}{$^1 P$} & \multicolumn{1}{c}{$^3 P$} \\
            \midrule
            $\delta_0$                     & -0.012\,141\,811\,35(4)    & 0.068\,355\,878\,73(4)     \\
            $\delta_2$                     & 0.007\,519\,166(9)         & -0.018\,631\,209(10)       \\
            $\delta_4$                     & 0.013\,9753(5)             & -0.012\,324\,3(6)          \\
            $\delta_6$                     & 0.004\,867(13)             & -0.008\,066(15)            \\
            $\delta_8$                     & 0.001\,13(10)              & -0.004\,91(12)             \\
            \bottomrule
        \end{tabular}
    \end{threeparttable}
\end{table}

\clearpage

\end{document}